\documentclass[sn-mathphys]{sn-jnl}

\usepackage{mathtools}
\usepackage{tensor}
\usepackage{subcaption}
\usepackage{IEEEtrantools}
\usepackage{bbold}

\jyear{2021}%

\theoremstyle{thmstyleone}%
\theoremstyle{thmstyletwo}%

\theoremstyle{thmstylethree}%

\usepackage[scaled=1.15,mathscr]{urwchancal}

\begin{document}

\title[\textcolor{white}{}]{On-shell Recursion Relations for Tree-level Closed String Amplitudes}


\author{\fnm{Pongwit} \sur{Srisangyingcharoen}}\email{pongwits@nu.ac.th}
\author{\fnm{Aphiwat} \sur{Yuenyong}}\email{aphiwaty64@nu.ac.th}

\affil{\orgdiv{The Institute for Fundamental Study}, \orgname{Naresuan University}, \orgaddress{\city{Phisanulok}, \postcode{65000}, \country{Thailand}}}


\abstract{We derive a general expression for on-shell recursion relations of closed string tree-level amplitudes. Starting with the string amplitudes written in the form of the Koba-Nielsen integral, we apply the BCFW shift to deform them. In contrast to open string amplitudes, where poles are explicitly determined by the integration over vertex positions, we utilize Schwinger's parametrization to handle the pole structure in closed strings. Our analysis reveals that the shifted amplitudes contain $\delta$-function poles, which yield simple poles upon taking residues. This allows us to present a general expression for the on-shell recursion relation for closed strings. Additionally, we offer an alternative method for computing the residue of the shifted amplitudes by factorizing an $n$-point closed string amplitude into two lower-point amplitudes. This is achieved by inserting a completeness relation that includes all possible closed string states in the Fock space. Our results are consistent with those previously obtained.}

\keywords{Bosonic string, On-shell recursion relations}



\maketitle

\section{Introduction}\label{sec1}
Since the beginning of the string theory, understanding string scattering amplitudes has been a fundamental concern among string theorists. It was long known that in the limit of low energies, amplitudes in string theory reproduce those in QFT such as Yang-Mills \cite{Neveu:1971mu, Green:1982sw} and Einstein theory \cite{Scherk:1974ca, Yoneya:1974jg} plus $\alpha'$ corrections \cite{TSEYTLIN1986391, Gross:1986iv, Koerber:2001uu, Metsaev:1986yb, Bergshoeff:1989de, Garousi:2019mca}. The connection between string theory and QFT provides useful applications in both theories. Understanding the structure of string amplitudes would provide a better insight into those of quantum field theories. Concrete examples are the celebrated Kawai-Lewellen-Tye (KLT) relations \cite{Kawai:1985xq} which relates closed string amplitudes in terms of products of two open string amplitudes giving alternative descriptions of gravity as the square of gauge theory. These non-linear relations were proven later in the context of QFT \cite{Bjerrum-Bohr:2010kyi, Bjerrum-Bohr:2010mtb}.

 Another interesting structure was discovered by Plahte \cite{Plahte:1970wy} which are linear relations among color-ordered open string scattering amplitudes. These are currently known as monodromy relations. In the field theory limit, the relations reduce to the BCJ relations of Bern, Carrasco and Johansson \cite{Bern:2008qj} and the Kleiss-Kujif relations \cite{Kleiss:1988ne}. This results in a reduction of the number of color-ordered amplitudes from $(n-1)!$ as given by a cyclic property of the trace down to $(n- 3)!$ \cite{Bjerrum-Bohr:2009ulz, Stieberger:2009hq}. The monodromy relations among partial open string amplitudes can be captured by polygons in the complex plane \cite{Srisangyingcharoen:2020lhx}. Much research has explored structural relations between string scattering amplitudes \cite{Srisangyingcharoen:2020lhx, Stieberger:2023nol, Yuenyong:2024ebe, Gerken:2020xfv, Stieberger:2016lng, Dorigoni:2022npe, Berkovits:2022ivl}.

During the early 2000s, advancements in the study of scattering amplitudes were notably influenced by the discovery of the Britto-Cachazo-Feng-Witten (BCFW) on-shell recursion relations \cite{Britto_2005,Britto2005NewRR}. These relations enabled the expression of tree-level amplitudes as products involving amplitudes with fewer particles. The key idea for deriving the on-shell recursion relations is based on the fact that any tree-level scattering amplitude is a rational function of the external momenta, thus, one can turn an amplitude $A_n$ into a complex meromorphic function $A_n(z)$ by deforming the external momenta through introducing a complex variable $z$. These deformed momenta, satisfying momentum conservation, are required to remain on-shell. For a scattering process involving $n$ particles, the selection of an arbitrary pair of particle momenta for shifting is permissible. Our choice is given by
\begin{subequations} \label{BCFW shift}
\begin{align}
    k_1 \rightarrow \hat{k}_1(z) =& k_1-qz \\
    k_n \rightarrow \hat{k}_n(z) =& k_n+qz
\end{align}
\label{shifted momenta2}
\end{subequations}
where $q$ is a reference momentum which obeys $q\cdot q=k_1\cdot q=k_n\cdot q=0$. 

The unshifted amplitude $A_n(z=0)$ can be obtained from a contour integration in which the contour is large enough to enclose all finite poles. According to the Cauchy's theorem,
\begin{equation}
    A_n(0)=\frac{1}{2\pi i}\oint dz \frac{A_n(z)}{z}-\sum_\text{poles} \text{Res}_{z=z_\text{poles}}\left(  \frac{A_n(z)}{z} \right), \label{cauchy}
\end{equation}
the unshifted amplitude at $z=0$ is equal to the sum of the residues over all the finite poles if the amplitude is well-behaved at large $z$ (which is the case for  most theories). For Yang-Mills theory, the residue at a finite pole is the product of two fewer-point amplitudes with an on-shell exchanged particle. In Yang-Mills a sum over the helicities of the intermediate gauge boson and in general theories a sum over all allowed intermediate particle states must also be done. In the general case, the BCFW recursion relation is 
\begin{equation}
    A_n(0)=\sum_{\substack{\text{poles}\\ \alpha}}\sum_{\substack{\text{physical}\\ \text{states}}} A_L(\dots,P(z_\alpha))\frac{2}{P^2+M^2}A_R(-P(z_\alpha),\dots)
    \label{BCFW}
\end{equation}
with $P$ being the momentum of the exchanged particle with mass $M$. 

The validity of equation (\ref{cauchy}) requires the absence of a pole at infinity. In the case that there exists such a pole, one must include the residue at infinity. However, the residue at this pole does not have a similar physical interpretation to the residues at finite poles. A detailed discussion can be found in \cite{Feng_2010}.

This paper aims to present expressions for on-shell recursion relations concerning closed string amplitudes. While existing literature has explored relations for open strings \cite{RutgerBoels_2008, Boels:2010bv, Chang:2012qs, Cheung:2010vn, Srisangyingcharoen:2024qyx}, the context of closed strings remains relatively unexplored. Herein, we seek to derive explicit expressions for the on-shell recursion relations of closed strings. 

Our main result is the derivation of a general framework for constructing closed string amplitudes through on-shell recursion relations. Unlike the case of field theory, where a finite set of intermediate particles is sufficient to generate higher-point amplitudes, closed string theory involves an infinite tower of string state excitations. This intrinsic feature renders it conceptually impossible to construct higher-point amplitudes using only lower-point amplitudes with the same external states. For instance, pure tachyonic amplitudes cannot be derived from lower-point pure tachyonic amplitudes alone, as is typical in field theory. Instead, they necessarily involve scattering amplitudes of tachyons and arbitrary string states.

In the next section, we explore how to obtain the on-shell recursion relations for closed string amplitudes, focusing on tachyonic amplitudes for simplicity (specifically, 4- and 5-point amplitudes). In Section 3, we generalize the results to amplitudes with arbitrary points and external string states. Section 4 verifies the general results through explicit computation of residues using the unitarity of the amplitudes. Finally, we conclude our findings in Section 5.


\section{On-shell Recursion Relations for Closed String Amplitudes at Fewer Points}\label{sec3}
\subsection{Four-Point Amplitudes}
In this section, we would like to formulate the on-shell recursion relations for the closed string amplitudes. For simplicity, we start our investigation with the simplest four-point tachyonic amplitude whose expression is
\begin{equation}
    \mathcal{A}^{\text{cl}}_4=\int d^2z \vert z\vert^{\alpha'p_1\cdot p_2}\vert 1-z\vert^{\alpha'p_2\cdot p_3} \label{4pt closed amp1}
\end{equation}
where $p_i$ are closed string momenta. We then choose to deform the momenta as
\begin{subequations} \label{closed BCFW shift}
\begin{align}
    p_1 \rightarrow \hat{p}_1(z) =& p_1-qz \\
    p_n \rightarrow \hat{p}_n(z) =& p_n+qz.
\end{align}
\label{shifted momenta}
\end{subequations}
For the shifted momenta to be on-shell, the reference momentum $q$ satisfies $q\cdot q=p_1\cdot q=p_n\cdot q=0$. The shifted amplitude takes the form
\begin{align}
    \mathcal{A}^{\text{cl}}_4(\hat{1},2,3,\hat{4})(\tilde{z})=\int d^2z  \vert z\vert^{\alpha'p_1\cdot p_2-2\tilde{z}}\vert 1-z\vert^{\alpha'p_2\cdot p_3}
\end{align}
where $\tilde{z}=\frac{\alpha'}{2}z(q\cdot p_2)$. Although closed string amplitudes lack a notion of color-ordering, we adopt a similar convention to track string states systematically. The amplitude can also be expressed symmetrically as:
\begin{equation}
    \mathcal{A}^{\text{cl}}_4(\tilde{z})=\frac{1}{4!}\sum_{\sigma} A_4^{\text{cl}}(\sigma(1,2,3,4))(\tilde{z}). \label{4pt closed amp2}
\end{equation}
where the sum is over all permutations $\sigma$  of the external momenta. This permutation simply relabels the momenta of the particles, with the first and last indices being the shifted ones. Although the subamplitudes are equal, their distinct integral forms allow us to access different pole channels, which is why they are expressed in this manner.

Another reason for expressing the amplitude in this manner is to ensure that the resulting expressions exhibit the expected pole structure in the $s-,t-,$ and $u-$channels, as dictated by the Virasoro-Shapiro formula. Remind that when $\tilde{z}=0$, the amplitude returns to the original one.

Let us first evaluate the integral $A^{\text{cl}}_4(\hat{1},2,3,\hat{4})(\tilde{z})$. It is not hard to see that there are possible poles regarding $\tilde{z}$ arising from the term $\vert z\vert^{\alpha'p_1\cdot p_2-2\tilde{z}}$. For this reason, we apply 
\begin{equation}
    \vert z\vert^{2a-2}=\frac{1}{\Gamma(1-a)}\int_0^\infty dt\ t^{-a} e^{-\vert z \vert^2t} \label{Schwinger}
\end{equation}
to rewrite
\begin{equation}
    A^{\text{cl}}_4(\hat{1},2,3,\hat{4})(\tilde{z})=\frac{1}{\Gamma(1-\beta)}\int d^2z \int_0^\infty dt\ t^{-\beta}e^{-\vert z \vert^2t}\vert 1-z\vert^{\alpha'p_2\cdot p_3} \label{closed a4-1}
\end{equation}
where
\begin{equation}
    \beta=\frac{\alpha'}{2}p_1\cdot p_2-\tilde{z}+1=\frac{\alpha'}{4}(p_1+p_2)^2-\tilde{z}-1.
\end{equation}
According to the binomial expansion, one can write
\begin{equation}
    \vert 1-z \vert^{\alpha'p_2\cdot p_3}=\sum_{a,b=0}^\infty \binom{\frac{\alpha'}{2}p_2\cdot p_3}{a}\binom{\frac{\alpha'}{2}p_2\cdot p_3}{b}(-1)^{a+b}z^a \bar{z}^b.
\end{equation}
This gives the complex Gaussian integral 
\begin{align}
    \int d^2z \ z^a \bar{z}^b e^{-\vert z \vert^2t}&=\Big(\frac{d}{d\bar{J}}\Big)^a\Big(\frac{d}{dJ}\Big)^b \int d^2z \ e^{-\vert z \vert ^2t+\bar{J}z+J\bar{z}}\bigg\vert_{J=\bar{J}=0} \nonumber \\
    &=\frac{2\pi i}{t}\Big(\frac{d}{d\bar{J}}\Big)^a\Big(\frac{d}{dJ}\Big)^b e^{\vert J \vert^2/t}\bigg\vert_{J=\bar{J}=0}. \label{complex gauss}
\end{align}
Only the case that $a=b$ would give non-zero contributions of the integration. Consequently, the amplitude (\ref{closed a4-1}) becomes
\begin{align}
    A^{\text{cl}}_4(\hat{1},2,3,\hat{4})(\tilde{z})=&\frac{2\pi i}{\Gamma(1-\beta)}\int_0^\infty dt \ t^{-\beta-1}\sum_{a=0}^\infty \binom{\frac{\alpha'}{2}p_2\cdot p_3}{a}^2 \frac{a!}{t^a} \label{A23}
\end{align}
which is troublesome to evaluate since it contains the divergent integral. To deal with this issue, we change the variable $t$ into $w$ by writing $t=e^w$, hence
\begin{equation}
   \int_0^\infty dt \ t^{-\beta-1-a}= \int_{-\infty}^\infty dw \ e^{-(\beta+a)w} \label{divergent}
\end{equation}
to which we can Wick rotate $w\rightarrow iw$ such that the integration over $w$ turns into the Dirac delta function, giving 
\begin{align}
    A^{\text{cl}}_4(\hat{1},2,3,\hat{4})(\tilde{z})=\frac{(2\pi i)^2}{\Gamma(1-\beta)}\sum_{a=0}^\infty \binom{\frac{\alpha'}{2}p_2\cdot p_3}{a}^2 a!\ \delta(\beta+a).
\end{align}
Following \cite{lepowsky2012introduction}, the formal distribution of the Dirac delta function reads
\begin{equation}
    \delta(z-w)=\sum_{n\in \mathbb{Z}} z^{-n-1}w^n=\sum_{n\in \mathbb{Z}} z^n w^{-n-1}.\label{dirac delta}
\end{equation}
Accordingly, one can write 
\begin{equation}
    \delta(\beta+a)=\sum_{n\in\mathbb{Z}}(-\beta)^{-n}\ a^{n-1}=\sum_{n\in\mathbb{Z}} \Big(\tilde{z}-\frac{\alpha'}{4}(p_1+p_2)^2)+1\Big)^{-n} \ a^{n-1}. \label{delta fn}
\end{equation}
The above summation provide poles of the order $n$  where $n\geq 1$ at $\tilde{z}=\frac{\alpha'}{4}(p_1+p_2)^2-1$. 

One obtains the unshifted $A^{\text{cl}}_4(\hat{1},2,3,\hat{4})(\tilde{z}=0)$ using the equation (\ref{cauchy}) which is
\begin{align}
    A^{\text{cl}}_4(\hat{1},2,3,\hat{4})(0)=- \text{Res}_{\tilde{z}=z^*}\Big( \frac{A^{\text{cl}}_4(\hat{1},2,3,\hat{4})(\tilde{z})}{\tilde{z}}\Big)
\end{align}
where $z^*=\frac{\alpha'}{4}(p_1+p_2)^2-1$. Note that the pole at infinity can be neglected, as the integrand $\vert z\vert^{\alpha'p_1\cdot p_2-2\tilde{z}}$ is suppressed when $\tilde{z}\rightarrow \infty$. Moreover, Some readers might find it surprising that a single pole, $\tilde{z}=z^*$, can generate an infinite number of poles corresponding to the particular scattering channels. As we will demonstrate shortly, this is achieved by transforming the infinitely many higher-order poles into an infinite set of simple poles, each evaluated at distinct points. To see this, we apply the formula
\begin{equation}
    \text{Res}_{z=z_0}f(z)=\frac{1}{(r-1)!}\frac{d^{r-1}}{dz^{r-1}}(z-z_0)^r f(z)\bigg\vert_{z=z_0}
\end{equation}
where the function $f(z)$ contains the pole of the order $r$ at $z=z_0$. Therefore, one can compute the residue
\begin{align}
    \text{Res}&_{\tilde{z}=z^*}\bigg( \frac{1}{\tilde{z}\Gamma(1-\beta)}\sum_{n=1}^\infty (\tilde{z}-z^*)^{-n} \ a^{n-1}\bigg) \nonumber \\
    &=\sum_{n=0}^\infty \frac{a^{n}}{n!}\bigg(\frac{d}{d\tilde{z}}\bigg)^n\bigg(\frac{1}{\tilde{z}\Gamma(1-\beta)}\bigg)\bigg\vert_{\tilde{z}=z^*} \label{residue}
\end{align}
Note that we discarded the summation over non-positive integers $n$ in the first line as they contain no poles. The negligence of the negative modes is helpful to disregard all singularities arising when $a$ equals zero. 

It turns out that the right-hand side of (\ref{residue}) is nothing but a Taylor expansion of the function $f(z^*+a)$ around the point $a=z^*$ where in our case $f(x)=1/(x\Gamma(1-\beta(x)))$. The residue (\ref{residue}) becomes
\begin{equation}
    \frac{1}{(z^*+a)\Gamma(1-\beta(z^*))}=\frac{1}{a!\left( \frac{\alpha'}{4}(p_1+p_2)^2+a-1\right)}.
\end{equation}
In general, we simply show that 
\begin{equation}
    \text{Res}_{z=z_0}\left(f(z)\delta(z-z_0) \right)=f(z_0). \label{residue delta}
\end{equation}

Consequently, the unshifted amplitude $A^{\text{cl}}_4(\hat{1},2,3,\hat{4})(\tilde{z}=0)$ takes the form
\begin{align}
        A^{\text{cl}}_4(\hat{1},2,3,\hat{4})(0)=\frac{4}{\alpha'}(2\pi)^2 \sum_{a=0}^\infty \binom{\frac{\alpha'}{2}p_2\cdot p_3}{a}^2 \frac{1 }{(p_1+p_2)^2+\frac{4}{\alpha'}(a-1)}. \label{BCFW closed1}
\end{align}
The denominators imply the mass spectrum of intermediate on-shell closed strings. Comparing the above expression with the recursion relations (\ref{BCFW}) suggests that the square of the binomial coefficients can be interpreted as a recursion involving products of three-point amplitudes. Specifically,
\begin{equation}
    A_L(\hat{1},2,\hat{P}(z_a))A_R(-\hat{P}(z_a),3,\hat{4})=\frac{(2\pi)^2}{\alpha'}\binom{\frac{\alpha'}{2}p_2\cdot p_3}{a}^2 \label{residue 4 pt}
\end{equation}
where $\hat{P}=-(\hat{p}_1+p_2)$ and the subamplitudes are evaluated at $z_a=\frac{(p_1+p_2)^2+4(a-1)/\alpha'}{2q\cdot p_2}$. A detailed computation of the residues of the relevant poles will be provided in Section 4.

Substitute (\ref{BCFW closed1}) to (\ref{4pt closed amp2}) gives the expression for on-shell recursion relation of four-point closed string amplitude as 
\begin{align}
    \mathcal{A}^{\text{cl}}_4=\frac{(8\pi)^2}{4!\alpha'} \sum_{a=0}^\infty  \Bigg(&\frac{1}{(p_1+p_2)^2+\frac{4}{\alpha'}(a-1)}\left[ \binom{\frac{\alpha'}{2}p_1\cdot p_3}{a}^2+\binom{\frac{\alpha'}{2}p_2\cdot p_3}{a}^2 \right]\nonumber \\ +&\frac{1 }{(p_1+p_3)^2 
    +\frac{4}{\alpha'}(a-1)}\left[ \binom{\frac{\alpha'}{2}p_1\cdot p_2}{a}^2+\binom{\frac{\alpha'}{2}p_2\cdot p_3}{a}^2 \right] \nonumber \\ +&\frac{1 }{(p_2+p_3)^2 
    +\frac{4}{\alpha'}(a-1)}\left[ \binom{\frac{\alpha'}{2}p_1\cdot p_2}{a}^2+\binom{\frac{\alpha'}{2}p_1\cdot p_3}{a}^2 \right]\Bigg)\label{BCFW 4 point}
\end{align}
Since there is no notion of color ordering in the closed string amplitudes, the momenta of intermediate closed string states are not necessarily a sum of adjacent momenta as in the case of open string amplitudes \cite{Srisangyingcharoen:2024qyx}. It is worth noting that the expression (\ref{BCFW 4 point}) exhibits a simple pole structure in the $s-,t-,$ and $u-$channels, similar to that of the Virasoro-Shapiro amplitude \cite{Virasoro:1969me, Shapiro:1970gy}. This is the primary motivation for symmetrizing the expression in (\ref{4pt closed amp2}).

\subsection{Five-Point Amplitudes}
Before moving to the discussion for $n$-point string scattering, we would like to consider another non-trivial case of five-point  tachyon scattering, allowing us to observe more patterns in the relations. Similar to the four-point amplitude, we write the full shifted amplitude as
\begin{equation}
    \mathcal{A}_5^{\text{cl}}(z)=\frac{1}{5!}\sum_{\sigma}A_5^{\text{cl}}(\sigma(1,2,3,4,5))(z) \label{A5}
\end{equation}
when the momenta are shifted using (\ref{shifted momenta}). Remember that we replace $\tilde{z}$ by $z$ for convenience. Again, the expression is summed over a $\sigma$ permutation of external momenta by which the first and the last indices are to be shifted. Let's consider the amplitude
\begin{align}
   A_5^{\text{cl}}(\hat{1},2,3,4,\hat{5})(z) = \int d^2z_2 d^2z_3 \ &\vert z_2 \vert^{\alpha' \hat{p}_1\cdot p_2} \vert z_3\vert^{\alpha' \hat{p}_1\cdot p_3} \vert1-z_2\vert^{\alpha' p_2\cdot p_4}  \nonumber \\
   \times & \vert 1-z_3\vert^{\alpha' p_3\cdot p_4} \vert z_2-z_3\vert^{\alpha' p_2 \cdot p_3}. \label{A234-1}
\end{align}
We then change the integral variables as follows:
\begin{equation}
    z_2=w_1w_2, \qquad z_3=w_2, \qquad \bar{z}_2=\bar{w}_1\bar{w}_2, \qquad \text{and} \quad \bar{z}_3=\bar{w}_2,
\end{equation}
to rewrite the expression (\ref{A234-1}) in the form
\begin{align}
 \int d^2 w_1 d^2 w_2 \,\, &\vert w_1\vert^{\alpha' p_1 \cdot p_2 - 2\tilde{z}_1} \vert w_2\vert^{\alpha' (p_1 \cdot p_2 + p_1 \cdot p_3+ p_2 \cdot p_3) -2\tilde{z}_2+2}  \nonumber \\ 
    \times& \vert 1-w_1\vert^{\alpha' p_2 \cdot p_3} \vert 1-w_2\vert^{\alpha'p_3 \cdot p_4} \vert 1-w_1w_2\vert^{\alpha' p_2 \cdot p_4} \label{A234-11}
\end{align}
where $\tilde{z}_j$ = $\frac{\alpha ' }{2}z q\cdot( \sum_{i=2}^{j+1}p_i)$ for $j=1,2$.  This implies that there are two poles generated from $\tilde{z}_j$. Consider the $\tilde{z}_1$ poles, we apply the binomial expansions
\begin{subequations} \label{binom5}
\begin{align}
    \vert 1-w_1\vert^{\alpha' p_2 \cdot p_3}  &= \sum_{b,c= 0}^{\infty} \binom{\frac{\alpha '}{2} p_2 \cdot p_3}{b}\binom{\frac{\alpha '}{2} p_2 \cdot p_3}{c} (-1)^{b+c} w_1^{b}\bar{w}_1^{c} \\
        \vert 1-w_1w_2\vert^{\alpha' p_2 \cdot p_4} &= \sum_{d,e = 0}^{\infty} \binom{\frac{\alpha '}{2} p_2 \cdot p_4}{d}\binom{\frac{\alpha '}{2} p_2 \cdot p_4}{e} (-1)^{d+e} (w_1w_2)^{d}(\bar{w}_1\bar{w}_2)^{e},
\end{align}
\end{subequations}
to rewrite the amplitude (\ref{A234-1}) in the form
\begin{align}
     A_5^{\text{cl}}(\hat{1},2,3,4,\hat{5})(z) =&\sum_{b,c,d,e=0}^\infty \binom{\frac{\alpha '}{2} p_2 \cdot p_3}{b}\binom{\frac{\alpha '}{2} p_2 \cdot p_3}{c}\binom{\frac{\alpha '}{2} p_2 \cdot p_4}{d}\binom{\frac{\alpha '}{2} p_2 \cdot p_4}{e} \nonumber\\ 
     &\times (-1)^{b+c+d+e}\left(\int d^2w_1 \int_0^\infty dt \, \frac{t^{-a_1}e^{-\vert w_1\vert^2 t}}{\Gamma(1-a_1)} w_1^{b+d} \bar{w}_1^{c+e} \right)\nonumber \\
     &\times \int d^2w_2 \vert w_2 \vert^{2a_2-2} w_2^d\bar{w}_2^e\vert 1-w_2\vert^{\alpha'p_3 \cdot p_4} \label{A234-2}
\end{align}
where 
\begin{align}
    a_1 &=\frac{\alpha'}{4}(p_1+p_2)^2-1-\tilde{z}_1  \\
    \intertext{and} a_2 &=\frac{\alpha'}{4}(p_1+p_2+p_3)^2-1-\tilde{z}_2. \label{a12}
\end{align}
Note that to obtain (\ref{A234-2}), the relation (\ref{Schwinger}) was utilized. Integrate out $w_1$ using the complex Gaussian integral (\ref{complex gauss}), the second line term in the parenthesis takes the form
\begin{equation}
   \frac{2\pi i}{\Gamma(1-a_1)} (b+d)!\delta_{b+d,c+e}\int_0^\infty dt \ t^{-a_1-b-d-1} 
\end{equation}
by which we can turn the integral to the Dirac delta function via Wick rotation, i.e.
\begin{equation}
   \frac{(2\pi i)^2}{\Gamma(1-a_1)} (b+d)!\delta_{b+d,c+e} \delta(a_1+b+d).
\end{equation}
Using (\ref{residue delta}), one can find
\begin{align}
   \sum_{\text{poles} z^*_1}&\text{Res}_{z=z^*_1}\left( \frac{A_5^{\text{cl}}(\hat{1},2,3,4,\hat{5})(z)}{z}\right)=(2\pi i)^2\sum_{\substack{b,c,d,e=0\\b+d=c+e}}^\infty \binom{\frac{\alpha '}{2} p_2 \cdot p_3}{b}\binom{\frac{\alpha '}{2} p_2 \cdot p_3}{c} \nonumber \\
    &\times\binom{\frac{\alpha '}{2} p_2 \cdot p_4}  {d}\nonumber\binom{\frac{\alpha '}{2} p_2 \cdot p_4}{e}\frac{2/\alpha' }{(p_1+p_2)^2+\frac{4}{\alpha'}(b+d-1)} \tilde{A}^{\text{cl}}_4(\hat{P}(z^*_1),3,4,\hat{5}) \\ \label{res5-z1}
\end{align}
where $z^*_1=\frac{2}{\alpha'q\cdot p_2}\left(\frac{\alpha'}{4}(p_1+p_2)^2+(b+d-1) \right)$. The integral in the third line of (\ref{A234-2}) is captured by the four-point deformed amplitude $A^{\text{cl}}_4$ given by
\begin{equation}
    \tilde{A}^{\text{cl}}_4(\hat{P}(z^*_1),3,4,\hat{5})=\int d^2w_2 \vert w_2 \vert^{\alpha'\hat{P}(z^*_1)\cdot p_3-2(b+d)}w_2^d\bar{w}_2^e\vert 1-w_2\vert^{\alpha'p_3 \cdot p_4} \label{4pt deform amp}
\end{equation}
where the intermediate momentum $\hat{P}=\hat{p}_1+p_2$. The propagator suggests the full spectrum of closed string states, i.e.
\begin{equation}
    (p_1+p_2)^2+\frac{4}{\alpha'}(b+d-1), \label{propagator 5 pt}
\end{equation}
whose integer $N=b+d$ is a level of the string state. Notice the condition for the summing variables $b+d=c+e=N$, which refers to a level-matching condition in a closed string state.

Recall that although we began with tachyon amplitudes where the exponents of the holomorphic and anti-holomorphic parts are identical, the fewer-point amplitudes obtained through the derived relation are not purely tachyonic. This is evident in the four-point amplitude (\ref{4pt deform amp}), where there is no constraint requiring the holomorphic and anti-holomorphic exponents \(d\) and \(e\) to be equal (albeit $b+d=c+e$). This is consistent with the understanding that fewer-point amplitudes must account for all intermediate string states associated with the propagator (\ref{propagator 5 pt}). 

Alternatively, the appearance of external states as different excitations in the physical string spectrum arises from the unitarity of the amplitudes. This unitarity allows higher-point amplitudes to be factorized into products of fewer-point amplitudes through the insertion of a unity operator. By employing a complete set of string states in Fock space as the basis, the fewer-point amplitudes naturally incorporate various external string state excitations. A more detailed discussion is provided in Section 4.

Moving to the $\tilde{z}_2$-poles, we follow the same steps by binomial expanding the terms
\begin{equation}
    \vert 1-w_2\vert^{\alpha' p_3 \cdot p_4} \qquad \text{and} \qquad \vert 1-w_1w_2\vert^{\alpha' p_2 \cdot p_4}
\end{equation}
from the expression (\ref{A234-11}) to obtain 
\begin{align}
    A_5^{\text{cl}}(\hat{1},2,3,4,\hat{5})(z) =&\sum_{b,c,d,e=0}^\infty \binom{\frac{\alpha '}{2} p_3 \cdot p_4}{b}\binom{\frac{\alpha '}{2} p_3 \cdot p_4}{c}\binom{\frac{\alpha '}{2} p_2 \cdot p_4}{d}\binom{\frac{\alpha '}{2} p_2 \cdot p_4}{e} \nonumber\\ 
     &\times (-1)^{b+c+d+e}\left(\int d^2w_2 \int_0^\infty dt \, \frac{t^{-a_2}e^{-\vert w_2\vert^2 t}}{\Gamma(1-a_2)} w_2^{b+d} \bar{w}_2^{c+e} \right)\nonumber \\
     &\times \int d^2w_1 \vert w_1 \vert^{2a_1-2} w_1^d \bar{w}_1^e\vert 1-w_1\vert^{\alpha'p_2 \cdot p_3} \label{A234-3}
\end{align}
where the relation (\ref{Schwinger}) was utilized. The parameters $a_1$ and $a_2$ were introduced in (\ref{a12}). It is not hard to repeat a calculation similar to the one presented earlier by which one obtains the residues at $\tilde{z}_2$-poles
\begin{align}
   &\sum_{\text{poles} z^*_2}\text{Res}_{z=z^*_2}\left( \frac{A_5^{\text{cl}}(\hat{1},2,3,4,\hat{5})(z)}{z}\right)=(2\pi i)^2\sum_{\substack{b,c,d,e=0\\b+d=c+e}}^\infty \binom{\frac{\alpha '}{2} p_3 \cdot p_4}{b}\binom{\frac{\alpha '}{2} p_3 \cdot p_4}{c} \nonumber \\
    &\times\binom{\frac{\alpha '}{2} p_2 \cdot p_4}  {d}\nonumber\binom{\frac{\alpha '}{2} p_2 \cdot p_4}{e}\frac{2/\alpha' }{(p_1+p_2+p_3)^2+\frac{4}{\alpha'}(b+d-1)} \tilde{A}^{\text{cl}}_4(\hat{1},2,3,\hat{P}'(z^*_2)) \\ \label{res5-z2}
\end{align}
where now $z^*_2=\frac{2}{\alpha'q\cdot (p_2+p_3)}\left(\frac{\alpha'}{4}(p_1+p_2+p_3)^2+(b+d-1) \right)$ and 
\begin{equation}
    \tilde{A}^{\text{cl}}_4(\hat{1},2,3,\hat{P}'(z^*_2))=\int d^2w_1 \vert w_1 \vert^{\alpha'\hat{p}_1(z^*_2)\cdot p_3}w_1^d\bar{w}_1^e\vert 1-w_1\vert^{\alpha'p_2 \cdot p_3}.
\end{equation}
with $\hat{P}'=-\hat{p}_1-p_2-p_3$.

To sum up, the on-shell recursion relation for $A_5^{\text{cl}}(\hat{1},2,3,4,\hat{5})(z=0)$ reads
\begin{align}
    A_5^{\text{cl}}(1,2,3,4,5)=\frac{(2\pi)^2}{\alpha'}\Bigg( &\sum_{\substack{b,c,d,e=0\\b+d=c+e}}^\infty \frac{\mathcal{R}_{\{b,c,d,e\}}}{(p_1+p_2)^2+\frac{4}{\alpha'}(b+d-1)} \nonumber \\
    +&\sum_{\substack{b',c',d',e'=0\\b'+d'=c'+e'}}^\infty \frac{\mathcal{S}_{\{b',c',d',e'\}}}{(p_1+p_2+p_3)^2+\frac{4}{\alpha'}(b'+d'-1)} \Bigg) \label{BCFW 5 point}
\end{align}
where
$\mathcal{R}_{\{b,c,d,e\}}$ and $\mathcal{S}_{\{b,c,d,e\}}$ can be identified with the right-hand side of the equation (\ref{res5-z1}) and (\ref{res5-z2}) respectively. Remind that to include all possible poles, one needs to sum up the remaining sub-amplitudes in (\ref{A5}) with $z=0$.

Before ending this section, let us comment on a choice independence of the reference momentum $q$. Although the expression (\ref{BCFW 5 point}) includes $q-$dependent subamplitudes arising from the shift (\ref{closed BCFW shift}), the sum of all poles should, in principle, reconstruct the original amplitude $A(z=0)$, which is independent of the choice of $q$. While this independence is evident in the case of the four-point recursion relations (\ref{BCFW 4 point}), it is elusive for the 5-point expression.


\section{On-shell Recursion Relations for $n$-point Closed String Amplitudes}

Now we would like to investigate the closed string amplitudes in a more general setting. Consider the general expression for tree-level closed string amplitude which is
\begin{align}
    \mathcal{A}^{\text{cl}}_n=\int_{\mathbb{C}}\prod_{i=1}^n d^2 z_i \frac{\vert z_{ab}\vert^2\vert z_{ac}\vert^2\vert z_{bc} \vert^2}{d^2z_a d^2z_b d^2z_c} \prod_{1\leq i<j\leq n}\vert z_i-z_j\vert^{\alpha'p_i\cdot p_j}\mathcal{F}_n\widetilde{\mathcal{F}}_n 
\end{align}
where $\vert z_{ij}\vert^2=\vert z_i-z_j\vert^2$. The position $z_a$, $z_b$ and $z_c$ are to be fixed to any distinct points in the complex plane. A traditional choice is to set them to be $z_a=z_1=0$, $z_b=z_{n-1}=1$ and $z_c=z_n=\infty$.

The function $\mathcal{F}_n$ and $\widetilde{\mathcal{F}}_n$ are external state-dependent functions of holomorphic and anti-holomorphic parts respectively. For closed string tachyons, $\mathcal{F}_n=\widetilde{\mathcal{F}}_n=1$. For the first-excited closed bosonic strings,
\begin{align}
    \mathcal{F}_n\widetilde{\mathcal{F}}_n=\exp \bigg[&\bigg( \sum_{i>j}\frac{\xi_i\cdot \xi_j}{(z_i-z_j)^2}+\frac{\bar\xi_i\cdot \bar\xi_j}{(\bar{z}_i-\bar{z}_j)^2}\bigg)\nonumber \\
    &+\sqrt{\alpha'}\sum_{i\neq j}\bigg( \frac{p_i\cdot \xi_j}{(z_i-z_j)} +\frac{p_i\cdot \bar\xi_j}{(\bar{z}_i-\bar{z}_j)} \bigg)\bigg]\bigg\vert_{\text{multilinear in } \xi,\bar\xi}
\end{align}
when the polarizations for the external states $\xi^{\mu\nu}_i=\xi^\mu_i\bar{\xi}^\nu_i$ are identified. In general, the closed string amplitude can be captured by the complex integral
\begin{equation}
    \int_{\mathbb{C}}\prod_{i=2}^{n-2}d^2z_i \ \prod_{1\leq i<j\leq n} \ (z_i-z_j)^{\frac{\alpha'}{2}p_i\cdot p_j+\tilde{n}_{ij}}(\bar{z}_i-\bar{z}_j)^{\frac{\alpha'}{2}p_i\cdot p_j+\tilde{\bar{n}}_{ij}} \label{closed integral}
\end{equation}
with integers $\tilde{n}_{ij}$ and $\tilde{\bar{n}}_{ij}$ coming from the functions $\mathcal{F}_n$ and $\widetilde{\mathcal{F}}_n$ respectively.

 We then  applied a change of integral variables from $(z_i,\bar{z}_i)$ to $(w_i,\bar{w}_i)$ using
\begin{equation}
    z_i=\prod_{j=i-1}^{n-3} w_j \qquad \text{and} \quad \bar{z}_i=\prod_{j=i-1}^{n-3} \bar{w}_j
\end{equation}
for $i=2,3,\ldots, n-2$. The integral (\ref{closed integral}) takes the form
\begin{align}
    \int_{\mathbb{C}} \prod_{i=1}^{n-3} d^2w_i & \prod_{j=1}^{n-3} w_j^{s_{12\ldots j+1}+n_j} \prod_{l=j}^{n-3}\left( 1-\prod_{k=j}^l w_k \right)^{s_{j+1,l+2}+n_{j+1,l+2}} \nonumber \\
    \times&\prod_{j=1}^{n-3} \bar{w}_j^{s_{12\ldots j+1}+\bar{n}_j} \prod_{l=j}^{n-3}\left( 1-\prod_{k=j}^l \bar{w}_k \right)^{s_{j+1,l+2}+\bar{n}_{j+1,l+2}} \label{An}
\end{align}
where $s_{ij}=\frac{\alpha'}{4}(p_i+p_j)^2$ and $s_{12\ldots i}=\frac{\alpha'}{4}(p_1+p_2+\ldots+p_i)^2$. We will refer to the above integral as $A_n^{\text{cl}}(1,2,\ldots,n)$. Remind that we now express the $n-$point closed string amplitude as 
\begin{equation}
    \mathcal{A}_n^{\text{cl}}=\frac{1}{n!}\sum_{\sigma}A_n^{\text{cl}}(\sigma(1,2,3,\ldots,n)).\label{An-1}
\end{equation}
The exponents $n_i, \bar{n}_i, n_{ij}$ and $\bar{n}_{ij}$ are given as
\begin{align}
    n_{ij}&=\tilde{n}_{ij}+\frac{\alpha'}{4}(m_{i}^2+m_{j}^2)  && i\leq j \nonumber \\
    \bar{n}_{ij}&=\tilde{\bar{n}}_{ij}+\frac{\alpha'}{4}(m_{i}^2+m_{j}^2) && i\leq j \nonumber \\
    n_j&=j-1+\sum_{l=1}^{j}\sum_{k>l}^{j+1}\tilde{n}_{lk}+\frac{\alpha'}{4}\sum_{i=1}^{j+1}m_i^2, && 1\leq j\leq n-3 \nonumber \\ 
    \bar{n}_j&=j-1+\sum_{l=1}^{j}\sum_{k>l}^{j+1}\tilde{\bar{n}}_{lk}+\frac{\alpha'}{4}\sum_{i=1}^{j+1}m_i^2, && 1\leq j\leq n-3. \label{n-n}
\end{align}
Note that we have fixed $z_1, z_{n-1}$ and $z_n$ to $0, 1$ and $\infty$ respectively. It is worth noticing that $\tilde{n}_{ij}$ and $\tilde{\bar{n}}_{ij}$ are not totally independent to each other due to the level-matching condition, i.e.
\begin{equation}
    \sum_{i<j} \tilde{n}_{ij} = \sum_{i<j} \tilde{\bar{n}}_{ij}
\end{equation}
for all $i$.


When the momenta are shifted using (\ref{shifted momenta2}), the deformed amplitudes reads
\begin{align}
    A_n^{\text{cl}}&(\hat{1},2,\ldots,\hat{n})(z)=\int_{\mathbb{C}} \prod_{i=1}^{n-3} d^2w_i  \prod_{j=1}^{n-3} \vert w_j\vert^{2s_{12\ldots j+1}-2\tilde{z}_j} w_j^{n_j}\bar{w}_j^{\bar{n}_j} \nonumber \\
    \times &\prod_{j=1}^{n-3}\prod_{l=j}^{n-3}\left( 1-\prod_{k=j}^l w_k \right)^{s_{j+1,l+2}+n_{j+1,l+2}}\left( 1-\prod_{k=j}^l \bar{w}_k \right)^{s_{j+1,l+2}+\bar{n}_{j+1,l+2}} \label{deformed An}
\end{align}
where $\tilde{z}_i=\frac{\alpha'}{2}z (q\cdot \sum_{j=2}^{i+1} p_j)$. This implies that there are $n-3$ poles corresponding to $z_i$. Note that the remaining poles is from a permutation of external momenta suggested in (\ref{An-1}).

Analogy to the fewer-point cases, to evaluate the $z_i$-pole, we apply the Schwinger parametrization (\ref{Schwinger}) to the term
\begin{equation}
    \vert w_i\vert^{2s_{12\ldots i+1}-2\tilde{z}_i}=\frac{1}{\Gamma(1-a_i)}\int_0^\infty dt\ t^{-a_i} e^{-\vert w_i \vert^2t}
\end{equation}
where
\begin{equation}
    a_i=s_{12\ldots i+1}+1-\tilde{z}_i,
\end{equation}
together with applying binomial expansions to every term that contains $w_i$ and $\bar{w}_i$ in the product
\begin{equation}
     \prod_{j=1}^{n-3}\prod_{l=j}^{n-3}\left( 1-\prod_{k=j}^l w_k \right)^{s_{j+1,l+2}+n_{j+1,l+2}}\left( 1-\prod_{k=j}^l \bar{w}_k \right)^{s_{j+1,l+2}+\bar{n}_{j+1,l+2}} . \label{product polynomial}
\end{equation}
These expansions would generate $2(n-2-i)i$ summing indices. For convenience, we will assign the summing index $x_{jl}$ and $y_{jl}$ for $j\leq l$ when expanding the polynomial
\begin{align}
    &\left( 1-\prod_{k=j}^l w_k \right)^{S_{j+1,l+2}}\left( 1-\prod_{k=j}^l \bar{w}_k \right)^{\bar{S}_{j+1,l+2}}\nonumber \\
    &=\sum_{x_{jl}=0}^\infty\sum_{y_{jl}=0}^\infty \binom{S_{j+1,l+2}}{x_{jl}}\binom{\bar{S}_{j+1,l+2}}{y_{jl}}(-1)^{x_{jl}+y_{jl}} \left(\prod_{k=j}^l w_k\right)^{x_{jl}}\left(\prod_{k=j}^l \bar{w}_k\right)^{y_{jl}} \label{binomial exp}
\end{align}
where $S_{j+1,l+2}\equiv s_{j+1,l+2}+n_{j+1,l+2}$ and $\bar{S}_{j+1,l+2}\equiv s_{j+1,l+2}+\bar{n}_{j+1,l+2}$ for short. We will use the above expansion to the product (\ref{product polynomial}) for which $j<l$ and $i\in[j,l]$ (Note that $i$ refers to the index where the poles $z_i$ will be determined). This gives
\begin{align}
       \prod_{j<l}&\left( 1-\prod_{k=j}^l w_k \right)^{S_{j+1,l+2}}\left( 1-\prod_{k=j}^l \bar{w}_k \right)^{\bar{S}_{j+1,l+2}}= \nonumber \\
       &\prod_{\substack{j< l\\i \notin [j,l]}}\left( 1-\prod_{k=j}^l w_k \right)^{S_{j+1,l+2}} \prod_{\substack{j< l\\i \in [j,l]}}\left(  \sum_{x_{jl}=0}^\infty \binom{S_{j+1,l+2}}{x_{jl}} (-1)^{x_{jl}} \left(\prod_{k=j}^l w_k  \right)^{x_{jl}} \right) \nonumber \\
       \times&\prod_{\substack{j< l\\i \notin [j,l]}}\left( 1-\prod_{k=j}^l \bar{w}_k \right)^{\bar{S}_{j+1,l+2}} \prod_{\substack{j< l\\i \in [j,l]}}\left(  \sum_{y_{jl}=0}^\infty \binom{\bar{S}_{j+1,l+2}}{y_{jl}} (-1)^{y_{jl}} \left(\prod_{k=j}^l \bar{w}_k  \right)^{y_{jl}} \right).
\end{align}
Note that we only apply the expansions for the case where $i \in [j,l]$ as discussed. Accordingly, we can write (\ref{deformed An}) as
\begin{align}
    \int_{\mathbb{C}} &\left(\prod_{\substack{k=1\\k\neq i}}^{n-3} d^2w_k \right)\prod_{\substack{j=1\\j\neq i}}^{n-3} \vert w_j\vert^{2s_{12\ldots j+1}-2\tilde{z}_j} w_j^{n_j}\bar{w}_j^{\bar{n}_j} \nonumber \\
    \times &\prod_{\substack{j< l\\i \notin [j,l]}}\left( 1-\prod_{k=j}^l w_k \right)^{S_{j+1,l+2}}\left( 1-\prod_{k=j}^l \bar{w}_k \right)^{\bar{S}_{j+1,l+2}} \nonumber\\
    \times & \prod_{\substack{j< l\\i \in [j,l]}}\left( 
        \sum_{x_{jl}=0}^\infty \binom{S_{j+1,l+2}}{x_{jl}} (-1)^{x_{jl}} \left(\prod_{\substack{k=j\\k\neq i}}^l w_k  \right)^{x_{jl}} \right. \nonumber \\
        &\qquad\times \left. \sum_{y_{jl}=0}^\infty \binom{\bar{S}_{j+1,l+2}}{y_{jl}} (-1)^{y_{jl}} \left(\prod_{\substack{k=j\\k\neq i}}^l \bar{w}_k  \right)^{y_{jl}} \right) \nonumber \\
    \times & \int d^2w_i \int_0^\infty dt \, \frac{t^{-a_i}e^{-\vert w_i\vert^2 t}}{\Gamma(1-a_i)} w_i^{\sum_{j<l}x_{jl}+n_i} \bar{w}_i^{\sum_{j<l}y_{jl}+\bar{n}_i}. \label{deformed An-2}
\end{align}
We can employ the complex Gaussian integral (\ref{complex gauss}) to the last line of (\ref{deformed An-2}) giving
\begin{align}
    \frac{2\pi i \left(\sum_{j<l}x_{jl}+n_i\right)!}{\Gamma(1-a_i)}\int_0^\infty dt \ t^{-(a_i+\sum_{j<l}x_{jl}+n_i+1)} \label{poly integral}
\end{align}
together with the condition that 
\begin{equation}
    \sum_{j<l}x_{jl}+n_i=\sum_{j<l}y_{jl}+\bar{n}_i.
\end{equation}
Similar to the fewer-point amplitudes, we turn the integral (\ref{poly integral}) into Dirac delta function that is
\begin{equation}
    \int_0^\infty dt \ t^{-(a_i+\sum_{j<l}x_{jl}+n_i+1)} =2\pi i \delta(a_i+\sum_{j<l}x_{jl}+n_i).
\end{equation}
Using the formula (\ref{residue delta}), one is allowed to determine the residue
\begin{align}
    \sum_{\text{poles} z^*_i}&\text{Res}_{z=z^*_i}\left( \frac{A_n^{\text{cl}}(\hat{1},2,\ldots,\hat{n})(z)}{z}\right)
\end{align}
where $z^*_i=\frac{2}{\alpha'q\cdot(\sum_{j=2}^{i+1}p_j)}(s_{12\ldots i+1}+\sum_{j<l}x_{jl}+n_i+1)$ and $i$ runs from $1$ to $n-3$ which equals to
\begin{align}
-(2\pi)^2\sum_{\substack{x_{jl},y_{jl}=0\\ \sum x_{jl}+n_i=\sum y_{jl}+\bar{n}_i}}^\infty \left( \frac{\mathcal{B}_i(x_{jl},y_{jl},\{n\},z)\lvert_{z=z^*_i}}{s_{12\ldots i+1}+n_i+1+\sum_{j<l}x_{jl}} \right)
\end{align}
where
\begin{align}
    \mathcal{B}_i&(x_{jl},y_{jl},\{n\},z)=\int_{\mathbb{C}} \left(\prod_{\substack{k=1\\k\neq i}}^{n-3} d^2w_k \right)\prod_{\substack{j=1\\j\neq i}}^{n-3} \vert w_j\vert^{2s_{12\ldots j+1}-2\tilde{z}_j} w_j^{n_j}\bar{w}_j^{\bar{n}_j} \nonumber \\
    \times &\prod_{j=1}^{n-3}\prod_{\substack{j< l\\i \notin [j,l]}}\left( 1-\prod_{k=j}^l w_k \right)^{S_{j+1,l+2}}\left( 1-\prod_{k=j}^l \bar{w}_k \right)^{\bar{S}_{j+1,l+2}} \nonumber\\
    \times & \prod_{j=1}^{n-3}\prod_{\substack{j< l\\i \in [j,l]}}\left( 
         \binom{S_{j+1,l+2}}{x_{jl}} \binom{\bar{S}_{j+1,l+2}}{y_{jl}}(-1)^{x_{jl}+y_{jl}} \left(\prod_{\substack{k=j\\k\neq i}}^l w_k  \right)^{x_{jl}} \left(\prod_{\substack{k=j\\k\neq i}}^l \bar{w}_k  \right)^{y_{jl}} \right). \label{residue B}
\end{align}
Since the above residue is evaluated at the $\tilde{z}_i-$pole, one needs to include all the possible poles $\tilde{z}_i$ by summing over $i$. With this in mind, it is helpful to rename the summation variables $x_{jl}$ and $y_{jl}$ as $x^{\scriptscriptstyle{(i)}}_{jl}$ and $y^{\scriptscriptstyle{(i)}}_{jl}$ to clarify which poles they correspond to. Therefore, one obtains the unshifted sub-amplitude
\begin{align}
    A_n^{\text{cl}}(1,2,\ldots,n)=(2\pi)^2\sum_{i=1}^{n-3}\sum_{\substack{x^{\scriptscriptstyle{(i)}}_{jl},y^{\scriptscriptstyle{(i)}}_{jl}=0\\ \sum x^{\scriptscriptstyle{(i)}}_{jl}+n_i=\sum y^{\scriptscriptstyle{(i)}}_{jl}+\bar{n}_i}}^\infty \left( \frac{\mathcal{B}_i(x^{\scriptscriptstyle{(i)}}_{jl},y^{\scriptscriptstyle{(i)}}_{jl},\{n\},z)\lvert_{z=z^*_i}}{s_{12\ldots i+1}+n_i+1+\sum_{j<l}x_{jl}} \right). \label{onshell-closed}
\end{align}
However, to obtain the full on-shell recursion relations, one is required to permute the external momenta as stated in (\ref{An-1}).

Moreover, the existence of an on-shell propagator implies that the function (\ref{residue B}) can be split into two objects referring to the
amplitudes of lower points, which is
\begin{align}
    \mathcal{B}_i(x_{jl},y_{jl},\{n\},z)=&\prod_{\substack{j\leq l\\i \in [j,l]}}\binom{S_{j+1,l+2}}{x_{jl}} \binom{S_{j+1,l+2}}{y_{jl}}(-1)^{x_{jl}+y_{jl}} \nonumber \\
    &\times \mathcal{A}^L_{i+1}(x_{jl},y_{jl},\{n\},z)\mathcal{A}^R_{n-i}(x_{jl},y_{jl},\{n\},z) \label{Bi}
\end{align}
where
\begin{align}
    \mathcal{A}^L_{i+1}(&x_{jl},y_{jl},\{n\},z)=\int_\mathbb{C}\left(\prod_{k=1}^{i-1} d^2w_k\right) \prod_{\substack{j=1}}^{i-1} \vert w_j\vert^{2s_{12\ldots j+1}-2\tilde{z}_j} w_j^{n_j}\bar{w}_j^{\bar{n}_j} \nonumber \\
    &\times\prod_{j=1}^{i-1}\prod_{l=j+1}^{n-3}\left(\prod_{k=j}^{i-1}w_k^{x_{jl}}\prod_{k=j}^{i-1}\bar{w}_k^{y_{jl}} \right)\nonumber \\
    &\times\prod_{j=1}^{i-1} \prod_{l=j}^{i-1}\left( 1-\prod_{r=j}^l w_r \right)^{s_{j+1,l+2}+n_{j+1,l+2}}\left( 1-\prod_{r=j}^l \bar{w}_r \right)^{s_{j+1,l+2}+\bar{n}_{j+1,l+2}}
\end{align}
and
\begin{align}
    \mathcal{A}^R_{n-i}(&x_{jl},y_{jl},\{n\},z)=\int_\mathbb{C}\left(\prod_{k=i+1}^{n-3} d^2w_k\right) \prod_{\substack{j=i+1}}^{n-3} \vert w_j\vert^{2s_{12\ldots j+1}-2\tilde{z}_j} w_j^{n_j}\bar{w}_j^{\bar{n}_j} \nonumber \\
    &\times\prod_{j=i+1}^{n-3}\prod_{l=j+1}^{n-3}\left(\prod_{k=i+1}^{l}w_k^{x_{jl}}\prod_{k=i+1}^{l}\bar{w}_k^{y_{jl}} \right)\nonumber \\
    &\times\prod_{j=i+1}^{n-3} \prod_{l=j}^{n-3}\left( 1-\prod_{r=j}^l w_r \right)^{s_{j+1,l+2}+n_{j+1,l+2}}\left( 1-\prod_{r=j}^l \bar{w}_r \right)^{s_{j+1,l+2}+\bar{n}_{j+1,l+2}}.
\end{align}
The functions $\mathcal{A}^L_{i+1}(x_{jl},y_{jl},\{n\},z)$ and $ \mathcal{A}^R_{n-i}(x_{jl},y_{jl},\{n\},z)$ resemble closed string amplitudes with $i+1$ and $n-i$ respectively.

It is not hard to see that the denominators in the expression (\ref{onshell-closed}) imply the propagators of the on-shell intermediate string states.  For a specific example of $n$-tachyon scattering, $n_i=-2$. The denominators suggest the full spectrum of closed strings, i.e.
\begin{equation}
    (k_1+k_2+\ldots+k_{i+1})^2+\frac{4}{\alpha'}\left(\sum_{j<l}x^{\scriptscriptstyle{(i)}}_{jl}-1\right)
\end{equation}
where the non-negative integer $\sum_{j<l}x^{\scriptscriptstyle{(i)}}_{jl}\equiv N$ is a level of string spectrum.

\section{State-by-state Calculation of On-shell Factorization of Closed String Amplitudes}
In this section, we would like to provide an alternative computation for the residue $\mathcal{B}_i(x^{\scriptscriptstyle{(i)}}_{jl},y^{\scriptscriptstyle{(i)}}_{jl},\{n\},z^*_i)$ using unitarity of the closed string amplitudes. Mathematically, we would like to show that 
\begin{align}
    \sum_{\text{physical} \ \Phi}\int \frac{d^D k}{(2\pi)^D}&\langle\phi_1\vert \mathcal{V}_2(p_2)\ldots \mathcal{V}_{i}(p_{i})\vert \Phi;k\rangle \nonumber \\
    &\otimes \langle\Phi;k\vert \mathcal{V}_{i+1}(p_{i+1})\ldots \mathcal{V}_{n-1}(p_{n-1})\vert \phi_n\rangle \label{goall}
\end{align}
resembles the function $\mathcal{B}_{i'}(x^{\scriptscriptstyle{(i')}}_{jl},y^{\scriptscriptstyle{(i')}}_{jl},\{n\},z^*_{i'})$ where $i=i'+1$. Notice that we use different indices to keep track of the poles in the expressions. Specifically, While $i'$ in $\mathcal{B}_{i'}$ corresponds to the pole where where the intermediate propagator carries momentum $p_1+p_2+\ldots+p_{i'+1}$, the index $i$ in the expression (\ref{goall}) implies to those with momentum $p_1+p_2+\ldots+p_{i}$ instead.

The expression (\ref{goall}) is nothing but a factorization of $n-$point string amplitudes into two on-shell fewer-point amplitudes. The states $\vert \phi \rangle$ is a string state and $\mathcal{V}_i$ refers to closed string vertex operators taking a form
\begin{equation}
    \mathcal{V}_i(p_i)=\int d^2z_i \ V_i(p_i,z_i,\bar{z_i}).
\end{equation}
In general, we can write
\begin{equation}
    V_i(p_i,z_i,\bar{z_i})=:f(z_i,\bar{z}_i)e^{ip_i\cdot X(z_i,\bar{z}_i)}:
\end{equation}
where the polarizations of higher-order string states are encoded in the function $f$. When $f=1$, they correspond to tachyons. Additionally, we must account for the global conformal symmetry, which allows us to freely fix three of the $z_i$ and $\bar{z}_i$ coordinates at arbitrary points in the complex plane.

The main challenge is that it is hard to determine physical intermediate string states. Luckily, according to \cite{Chang:2012qs}, one can by-pass this problem by enlarging the sum over intermediate physical states to over intermediate complete Fock space states. To see this, let consider such this factorization in gauge theory, i.e.
\begin{equation}
    \sum_{\text{physical} \ \epsilon}A_L(\sigma_L,k)A_R(\sigma_R,-k)=\sum_{\text{physical} \ \epsilon}\tilde{A}^\mu_L(\sigma_L,k)\epsilon_\mu \epsilon_\nu\tilde{A}^\nu_R(\sigma_R,-k)
\end{equation}
where $\epsilon_\mu$ is a physical polarization of gauge particle. From a Ward identity, it is possible to include not only physical polartization states but non-physical ones too. Similar to the string theory, we have the famous no-ghost theorem which guarantee that  the sum over physical states are equal to that over all states in Fock space. This means we could turn the expression \label{goal} to
\begin{align}
    \sum_{\text{Fock} \ \Phi}\int \frac{d^D k}{(2\pi)^D}&\langle\phi_1\vert \mathcal{V}_2(p_2)\ldots \mathcal{V}_{i}(p_{i})\vert \Phi;k\rangle \nonumber \\
    &\otimes \langle\Phi;k\vert \mathcal{V}_{i+1}(p_{i+1})\ldots \mathcal{V}_{n-1}(p_{n-1})\vert \phi_n\rangle \label{goal2}
\end{align}
More discussion on no-ghost theorem can be found in \cite{Chang:2012qs}.

The Fock space spans all possible string states that are constructed by acting right/left-moving creation operators $\alpha^\mu_{-n}$ and $\tilde{\alpha}^\mu_{-n}$ on the ground state, i.e.
\begin{equation}
    \Big(\alpha^{\mu_1}_{-n_1}\alpha^{\mu_2}_{-n_2}\ldots\alpha^{\mu_i}_{-n_i}\Big) \Big(\tilde{\alpha}^{\nu_1}_{-m_1}\tilde{\alpha}^{\nu_2}_{-m_2}\ldots\tilde{\alpha}^{\nu_j}_{-m_j}\Big)\vert0;k\rangle
\end{equation}
where $n_i$ and $m_i$ are positive integers. Due to the level-matching condition, the states are subject to $\sum_i n_i=\sum_i m_i$. In this work, we introduce a normalized Fock state for a closed string as
\begin{align}
    \vert\{\mathcal{N},\widetilde{\mathcal{N}}\};k\rangle=&\prod_{(n,N_n) \in \{\mathcal{N}\}}\frac{\left(\alpha^{\mu_{n,1}}_{-n}\ldots\alpha^{\mu_{n,N_n}}_{-n} \right)}{\sqrt{N_n!n^{N_n}}} \nonumber \\
    \times&\prod_{(m,N_m) \in \{\widetilde{\mathcal{N}}\}}\frac{\left(\tilde{\alpha}^{\tilde{\mu}_{m,1}}_{-m}\ldots\alpha^{\tilde{\mu}_{m,N_m}}_{-m} \right)}{\sqrt{N_m!m^{N_m}}}\vert 0;k\rangle. \label{state N}
\end{align}
The set $\{ \mathcal{N}\}$ contains elements $(n,N_n)$ that identify which vibrational mode to excite and how many times they operate respectively. The level of string spectrum $N$ is
\begin{equation}
    N=\sum_n n\times N_n.
\end{equation}
For example, in case $N=7$, one of the possible set $\{ \mathcal{N}\}$ is $\{(1,3),(2,2)\}$ meaning that when $n=1, N_1=3$ and $n=2, N_2=2$. Note that there are other possible combinations of $(n,N_n)$ that gives the set $\{\mathcal{N}\}$ for $N=7$ such as \{(1,7)\},  \{(1,1),(2,1),(4,1)\}, etc. The different set $\{\mathcal{N}\}$ provides different excited string states with the same level $N$. Moreover, the level-matching condition also requires $N=\tilde{N}$, yet this does not necessarily mean $\{\mathcal{N}\}=\{\widetilde{\mathcal{N}}\}$. For instance, one could have $\{\mathcal{N}\}=\{(1,3),(2,2)\}$ and $\{\widetilde{\mathcal{N}}\}=\{(3,1),(4,1)\}$ differently for the same $N=\tilde{N}=7$ that corresponds to state
\begin{equation}
    \left( \frac{\alpha_{-1}^{\mu_{1,1}}\alpha_{-1}^{\mu_{1,2}}\alpha_{-1}^{\mu_{1,3}}}{\sqrt{3!}}\right)\left( \frac{\alpha_{-2}^{\mu_{2,1}}\alpha_{-2}^{\mu_{2,2}}}{\sqrt{2!\times 4}}\right)\tilde{\alpha}_{-3}^{\tilde{\mu}_{3,1}}\tilde{\alpha}_{-4}^{\tilde{\mu}_{4,1}} \vert 0;k\rangle.
\end{equation}

Considering the states $\vert\{\mathcal{N},\widetilde{\mathcal{N}}\};k\rangle$ as a basis, we introduce a projection operator from the set of basis regarding to the center of mass momentum $k$ as
\begin{equation}
    \mathbb{P}(k)=\sum_{N=0}^\infty\sum_{\{\mathcal{N}\},\{\widetilde{\mathcal{N}}\}} \vert\{\mathcal{N},\widetilde{\mathcal{N}}\};k\rangle \mathcal{T}_{\{\mathcal{N},\widetilde{\mathcal{N}}\}} \langle \{\mathcal{N},\widetilde{\mathcal{N}}\};k\vert \label{projection}
\end{equation}
where 
\begin{align}
    \mathcal{T}_{\{\mathcal{N},\widetilde{\mathcal{N}}\}}=&\prod_{(n,N_n)\in\{\mathcal{N}\}}\left(g_{\mu_{n,1},\nu_{n,1}}\ldots g_{\mu_{n,N_n},\nu_{n,N_n}} \right) \nonumber \\
    \times& \prod_{(m,N_m)\in\{\widetilde{\mathcal{N}}\}}\left(g_{\tilde{\mu}_{m,1},\tilde{\nu}_{m,1}}\ldots g_{\tilde{\mu}_{m,N_m},\tilde{\nu}_{m,N_m}} \right). \label{T tensor}
\end{align}
For example, the first few terms of the projection operator reads
\begin{align}
    \mathbb{P}(k)=&\vert 0;k\rangle\langle 0;k\vert+\alpha_{-1}^{\mu_{1,1}}\tilde{\alpha}_{-1}^{\tilde{\mu}_{1,1}} \vert 0;k\rangle \left(g_{\mu_{1,1},\nu_{1,1}}g_{\tilde{\mu}_{1,1},\tilde{\nu}_{1,1}}\right)\langle 0;k\vert\alpha_{+1}^{\nu_{1,1}}\tilde{\alpha}_{+1}^{\tilde{\nu}_{1,1}} \nonumber \\
&+\left(\frac{\alpha_{-1}^{\mu_{1,1}}\alpha_{-1}^{\mu_{1,2}}}{\sqrt{2}}\right)\left(\frac{\tilde{\alpha}_{-1}^{\tilde{\mu}_{1,1}}\tilde{\alpha}_{-1}^{\tilde{\mu}_{1,2}}}{\sqrt{2}}\right)\vert 0;k\rangle \big(g_{\mu_{1,1},\nu_{1,1}}g_{\mu_{1,2},\nu_{1,2}}g_{\tilde{\mu}_{1,1},\tilde{\nu}_{1,1}}g_{\tilde{\mu}_{1,2},\tilde{\nu}_{1,2}}\big) \nonumber \\
&\langle 0;k\vert \left(\frac{\alpha_{+1}^{\nu_{1,1}}\alpha_{+1}^{\nu_{1,2}}}{\sqrt{2}}\right)\left(\frac{\tilde{\alpha}_{+1}^{\tilde{\nu}_{1,1}}\tilde{\alpha}_{+1}^{\tilde{\nu}_{1,2}}}{\sqrt{2}}\right) + \left(\frac{\alpha_{-1}^{\mu_{1,1}}\alpha_{-1}^{\mu_{1,2}}}{\sqrt{2}}\right)\tilde{\alpha}_{-2}^{\tilde{\mu}_{2,1}}\vert 0;k\rangle \nonumber \\
&\big( g_{\mu_{1,1},\nu_{1,1}}g_{\mu_{1,2},\nu_{1,2}}g_{\tilde{\mu}_{2,1},\tilde{\nu}_{2,1}} \big)\langle 0;k\vert \left(\frac{\alpha_{+1}^{\nu_{1,1}}\alpha_{+1}^{\nu_{1,2}}}{\sqrt{2}}\right)\tilde{\alpha}_{+2}^{\tilde{\nu}_{2,1}} +\ldots.
\end{align}
The operator obeys
\begin{equation}
    \mathbb{P}(k)\mathbb{P}(k')=(2\pi)^D\delta^{D}(k-k')\mathbb{P}(k).
\end{equation}
A completeness relation is recovered upon integrating over the momentum $k$, 
\begin{equation}
    \mathbb{1}=\int \frac{d^Dk}{(2\pi)^D}\ \mathbb{P}(k). \label{completeness}
\end{equation}

For simplicity, we will only focus our calculation on tachyonic closed string amplitudes whose non-integrated vertex operator is
\begin{equation}
    V(p_i,z_i,\bar{z}_i)=:e^{ip_i\cdot X(z_i,\bar{z}_i)}:
\end{equation}
where 
\begin{align}
X^\mu(z,\bar{z})=&X^\mu(z)+\bar{X}^\mu(\bar{z})\nonumber \\
=&x^\mu-i\frac{\alpha'}{2}p^\mu\ln{\vert z \vert^2}+i\sqrt{\frac{\alpha'}{2}}\sum_{n\neq 0}\frac{1}{n} \left( \alpha^\mu_nz^{-n}+\tilde{\alpha}^\mu_n\bar{z}^{-n}\right).
\end{align}
Accordingly, we can write
\begin{equation}
    V(p_i,z_i,\bar{z}_i)=e^{ip_i\cdot(x-i\frac{\alpha'}{2}p\ln\vert z \vert^2)}:\mathcal{W}(p_i):=\vert z\vert^{\alpha'p_i\cdot p-\frac{4}{\alpha'}}e^{ip_i\cdot x}:\mathcal{W}(p_i): \label{tachyon vertex}
\end{equation}
where \begin{equation}
    \mathcal{W}(p_i)=\exp\left( -\frac{\alpha'}{2}p_i\cdot \sum_{n\neq 0}\frac{1}{n} \left( \alpha^\mu_nz^{-n}+\tilde{\alpha}^\mu_n\bar{z}^{-n}\right)\right)
\end{equation}
and to obtain the above expression, the Baker–Campbell–Hausdorff formula was used.

\subsection{Residue of Tachyonic Four-Point Amplitude}
Let's first consider the four-point tachyon amplitude
\begin{align}
    \sum_{N=0}^\infty \Bigg(\sum_{\{\mathcal{N}\},\{\widetilde{\mathcal{N}}\}}&\int\frac{d^D k}{(2\pi)^D}\ \langle0;-p_1\vert V(p_2,z_2,\bar{z}_2) \vert\{\mathcal{N},\widetilde{\mathcal{N}}\};k\rangle \mathcal{T}_{\{\mathcal{N},\widetilde{\mathcal{N}}\}}  
    \nonumber \\
     &  \langle \{\mathcal{N},\widetilde{\mathcal{N}}\};k\vert V(p_3,z_3,\bar{z}_3)\vert 0;p_4\rangle  \Bigg)\equiv \sum_{N=0}^\infty \mathcal{I}_N
\end{align}
where we insert the completeness relation (\ref{completeness}) to split the amplitude into two three-point ones. The expression inside the parentheses is encapsulated as the function $\mathcal{I}_N$. Remind that the integrated vertex operators $\mathcal{V}(p_i)$ are changed into the non-integrated ones $V(p_i,z_i,\bar{z}_i)$ due to the $PSL(2,\mathbb{C})$ symmetry. Our choice is setting $z_2=\bar{z}_2=1$ and $z_3=\bar{z}_3=1$ to both sub-amplitudes. Therefore, 
\begin{align}
    \langle 0;-p_1&\vert V(p_2,z_2,\bar{z}_2) \vert\{\mathcal{N},\widetilde{\mathcal{N}}\};k\rangle= \langle0;-p_1\vert e^{ip_2\cdot x}:\mathcal{W}(p_2): \nonumber \\ \times &\prod_{(n,N_n) \in \{\mathcal{N}\}}\frac{\left(\alpha^{\mu_{n,1}}_{-n}\ldots\alpha^{\mu_{n,N_n}}_{-n} \right)}{\sqrt{N_n!n^{N_n}}}
    \prod_{(m,N_m) \in \{\widetilde{\mathcal{N}}\}}\frac{\left(\tilde{\alpha}^{\tilde{\mu}_{m,1}}_{-m}\ldots\alpha^{\tilde{\mu}_{m,N_m}}_{-m} \right)}{\sqrt{N_m!m^{N_m}}}\vert 0;k\rangle.
\end{align}
We then use the commutation relations
\begin{align}
    [\alpha^\mu_m,:\mathcal{W}(p_i):]&=\sqrt{\frac{\alpha'}{2}}p_i^\mu z^m :\mathcal{W}(p_i): \label{commutator} \\ \intertext{and} 
     [\tilde{\alpha}^\mu_m,:\mathcal{W}(p_i):]&=\sqrt{\frac{\alpha'}{2}}p_i^\mu \bar{z}^m :\mathcal{W}(p_i):
\end{align}
for $m\neq 0$, to obtain
\begin{align}
   (2\pi)^D\delta^{D}(p_1+p_2+k)&\prod_{(n,N_n) \in \{\mathcal{N}\}}\frac{\left(p_2^{\mu_{n,1}}\ldots p_2^{\mu_{n,N_n}} \right)}{\sqrt{N_n!n^{N_n}}}(-1)^{N_n}\left(\frac{\alpha'}{2}\right)^{N_n/2} \nonumber \\
   \times&\prod_{(m,N_m) \in \{\widetilde{\mathcal{N}}\}}\frac{\left(p_2^{\tilde{\mu}_{m,1}}\ldots p_2^{\tilde{\mu}_{m,N_m}} \right)}{\sqrt{N_m!m^{N_m}}}(-1)^{N_m}\left(\frac{\alpha'}{2}\right)^{N_m/2}.
\end{align}
Note that 
\begin{equation}
    \langle0;-p_1\vert e^{ip_2\cdot x}\vert 0;k\rangle=(2\pi)^D\delta^{D}(p_1+p_2+k)
\end{equation}
and we set $z_2=\bar{z}_2=1$. Similarly, we obtain
\begin{align}
    \langle \{\mathcal{N},\widetilde{\mathcal{N}}\};k\vert& V(p_3,z_3,\bar{z}_3)\vert 0;p_4\rangle=(2\pi)^D\delta^{D}(-k+p_3+p_4) \nonumber \\
    &\times \prod_{(n,N_n) \in \{\mathcal{N}\}}\frac{\left(p_2^{\nu_{n,1}}\ldots p_2^{\nu_{n,N_n}} \right)}{\sqrt{N_n!n^{N_n}}}\left(\frac{\alpha'}{2}\right)^{N_n/2} \nonumber \\
    &\times \prod_{(m,N_m) \in \{\widetilde{\mathcal{N}}\}}\frac{\left(p_2^{\tilde{\nu}_{m,1}}\ldots p_2^{\tilde{\nu}_{m,N_m}} \right)}{\sqrt{N_m!m^{N_m}}}\left(\frac{\alpha'}{2}\right)^{N_m/2}.
\end{align}
Consequently, this yields
\begin{align}
    \mathcal{I}_N=&(2\pi)^D\delta^D(p_1+p_2+p_3+p_4) \nonumber \\
    \times&\sum_{\{\mathcal{N}\},\{\widetilde{\mathcal{N}}\}}\left(\prod_{(n,N_n)\in\{\mathcal{N}\}}\frac{(-\frac{\alpha'}{2}p_2\cdot p_3)^{N_n}}{N_n!n^{N_n}}\prod_{(m,N_m)\in\{\widetilde{\mathcal{N}}\}}\frac{(-\frac{\alpha'}{2}p_2\cdot p_3)^{N_m}}{N_m!m^{N_m}}\right)  \label{IN 4-point}
\end{align}

In abstract algebra, the number 
\begin{equation}
    \prod_{i=1}^n \frac{n!}{i^{k_i}k_i!}
\end{equation}
where $n=\sum_{i=1}^n i\cdot k_i$ counts the number elements in a conjugacy class of the symmetric group $S_n$ with the same cycle structure. The number equals a non-negative (unsigned) version of the Stirling number of the first kind $\vert s(n,m)\vert$ that counts the number of permutations of $n$ objects having $m$ permutation cycles. In this case $m=\sum_{i=1}^n k_i$. The signed Stirling number of the first kind $s(n,m)$ relates to its absolute value $\vert s(n,m)\vert$ through
\begin{equation}
    s(n,m)=(-1)^{n-m}\vert s(n,m)\vert=(-1)^{n-m}\prod_{i=1}^n \frac{n!}{i^{k_i}k_i!}\label{Stirling}
\end{equation}
again with conditions 
\begin{equation}
    n=\sum_{i=1}^n i\cdot k_i \qquad \text{and} \qquad m=\sum_{i=1}^n k_i.
\end{equation}

This fact allows us to rewrite 
\begin{align}
    \sum_{\{\mathcal{N}\}}&\prod_{(n,N_n)\in\{\mathcal{N}\}}\frac{(-A)^{N_n}}{N_n!n^{N_n}}=\sum_{\{\mathcal{N}\}}(-A)^{\sum_{n} N_n}\prod_{(n,N_n)\in\{\mathcal{N}\}}\frac{1}{N_n!n^{N_n}}\nonumber \\
    &=\sum_{m=1}^N \frac{s(N,m)}{N!}(-1)^{N}A^{m}=(-1)^N\binom{A}{N} \label{binom}
\end{align}
with $m=\sum_n N_n$ and $\binom{A}{N}$ is an binomial coefficient. Using this connection, we can write (\ref{IN 4-point}) as
\begin{equation}
    \mathcal{I}_N=(2\pi)^D\delta^{D}\left(\sum_i p_i \right)\binom{\frac{\alpha'}{2}p_2\cdot p_3}{N}^2
\end{equation}
that exactly matches the expression (\ref{residue 4 pt}) up to a constant.

\subsection{Residue of Tachyonic General-Point Amplitudes}

Next, we would like to extend the calculation to the $n$-point amplitudes. Of course, we will only consider that all external states are tachyonic for simplicity. Our aim is to evaluate the sub-amplitude
\begin{align}
   \mathcal{I}_N= \sum_{\{\mathcal{N}\},\{\widetilde{\mathcal{N}}\}}&\int\frac{d^D k}{(2\pi)^D}\ \langle0;-p_1\vert \mathcal{V}(p_2)\ldots \mathcal{V}(p_{i}) \vert\{\mathcal{N},\widetilde{\mathcal{N}}\};k\rangle \mathcal{T}_{\{\mathcal{N},\widetilde{\mathcal{N}}\}}  
    \nonumber \\
     &  \langle \{\mathcal{N},\widetilde{\mathcal{N}}\};k\vert \mathcal{V}(p_{i+1})\ldots\mathcal{V}(p_{n-1})\vert 0;p_n\rangle.  \label{n-point correlation}
\end{align}

It is useful to consider the term
\begin{align}
    \langle0;-p_1\vert V(p_2,z_2,\bar{z}_2)\ldots V(p_{i},z_{i},\bar{z}_{i}) \alpha^\mu_{-m}\vert 0;k\rangle 
\end{align}
which equals 
\begin{equation}
        -\sqrt{\frac{\alpha'}{2}}\left( p^\mu_2z^m_2+ p^\mu_3z^m_3+\ldots+ p^\mu_{i}z^m_i\right)\langle0;-p_1\vert V(p_2,z_2,\bar{z}_2)\ldots V(p_{i},z_i,\bar{z}_i) \vert 0;k\rangle.
\end{equation}
The proof is straightforward using the commutation relation (\ref{commutator}). Remind that the expression (\ref{n-point correlation}) as closed string amplitude is incomplete without integrating vertex positions $d^2z_i$ over the complex plane which can be fixed up to 3 points using the global conformal symmetry. Without much effort, one can compute
\begin{align}
    \langle0;&-p_1\vert V(p_2,z_2,\bar{z}_2)\ldots V(p_{i},z_{i},\bar{z}_{i})\vert\{\mathcal{N},\widetilde{\mathcal{N}}\};k\rangle=\nonumber \\
    &\prod_{(n,N_n)\in\{\mathcal{N}\}} \frac{\prod_{k=1}^{N_n}\left( -\sqrt{\frac{\alpha'}{2}}\sum_{j=2}^{i}p^{\mu_{n,k}}_j z^n_j\right)}{\sqrt{N_n! n^{N_n}}} \nonumber \\
    &\times\prod_{(m,N_m)\in\{\widetilde{\mathcal{N}}\}} \frac{\prod_{k=1}^{N_m}\left( -\sqrt{\frac{\alpha'}{2}}\sum_{j=2}^{i}p^{\tilde{\mu}_{m,k}}_j \bar{z}^m_j\right)}{\sqrt{N_m! n^{N_m}}} \nonumber \\
    &\times \langle0;-p_1\vert V(p_2,z_2,\bar{z}_2)\ldots V(p_{i},z_{i},\bar{z}_{i})\vert0;k\rangle.
\end{align}
where the state $\vert\{\mathcal{N},\widetilde{\mathcal{N}}\};k\rangle$ is given in (\ref{state N}). Similarly,
\begin{align}
    \langle\{\mathcal{N},\widetilde{\mathcal{N}}\};&k\vert V(p_{i+1},z_{i+1},\bar{z}_{i+1})\ldots V(p_{n-1},z_{n-1},\bar{z}_{n-1})\vert0;p_{n}\rangle=\nonumber \\
    &\prod_{(n,N_n)\in\{\mathcal{N}\}} \frac{\prod_{k=1}^{N_n}\left( \sqrt{\frac{\alpha'}{2}}\sum_{j=i+1}^{n-1}p^{\nu_{n,k}}_j z^n_j\right)}{\sqrt{N_n! n^{N_n}}}\nonumber \\
    &\times \prod_{(m,N_m)\in\{\widetilde{\mathcal{N}}\}} \frac{\prod_{k=1}^{N_n}\left( \sqrt{\frac{\alpha'}{2}}\sum_{j=i+1}^{n-1}p^{\tilde{\nu}_{m,k}}_j \bar{z}^m_j\right)}{\sqrt{N_m! n^{N_m}}} \nonumber \\
    &\times \langle0;k\vert V(p_{i+1},z_{i+1},\bar{z}_{i+1})\ldots V(p_{n-1},z_{n-1},\bar{z}_{n-1})\vert0;p_{n}\rangle.
\end{align}
The $PSL(2,\mathbb{C})$ symmetry allows us to set $z_{i}=\bar{z}_i=1$ together with $z_{n-1}=\bar{z}_{n-1}=1$. 

That being so, one can write
\begin{align}
    \mathcal{I}_N=&\sum_{\{\mathcal{N}\},\{\widetilde{\mathcal{N}}\}}\int_\mathbb{C}\left( \prod_{\substack{j=2 \\j\neq i}}^{n-2} d^2z_j \right)\int\frac{d^D k}{(2\pi)^D}\nonumber \\
    &\times \left[ \prod_{(n,N_n)\in\{\mathcal{N}\}} \frac{\prod_{k=1}^{N_n}\frac{\alpha'}{2}\left( -\sum_{j=2}^{i}p^{\mu_{n,k}}_j z^n_j\right)\left( \sum_{j=i+1}^{n-1}p^{\nu_{n,k}}_j z^n_j\right)}{N_n! n^{N_n}}\right. \mathcal{T}_{\{\mathcal{N},\widetilde{\mathcal{N}}\}} \nonumber \\
    &\phantom{\times\Bigg[} \left.\prod_{(m,N_m)\in\{\widetilde{\mathcal{N}}\}} \frac{\prod_{k=1}^{N_m}\frac{\alpha'}{2}\left( -\sum_{j=2}^{i}p^{\tilde{\mu}_{m,k}}_j \bar{z}^m_j\right)\left( \sum_{j=i+1}^{n-1}p^{\tilde{\nu}_{m,k}}_j \bar{z}^m_j\right)}{N_m! m^{N_m}} \right] \nonumber \\
    &\times \langle0;-p_1\vert V(p_2,z_2,\bar{z}_2)\ldots V(p_{i},z_{i},\bar{z}_{i})\vert0;k\rangle \nonumber \\
    &\times \langle0;k\vert V(p_{i+1},z_{i+1},\bar{z}_{i+1})\ldots V(p_{n-1},z_{n-1},\bar{z}_{n-1})\vert0;p_{n}\rangle. \label{IN-1}
\end{align}
Keep in mind that now $z_i=\bar{z}_i=z_{n-1}=\bar{z}_{n-1}=1$. After performing tensorial contractions using $\mathcal{T}_{\{\mathcal{N},\widetilde{\mathcal{N}}\}}$ given in (\ref{T tensor}), the terms inside the squared bracket reads
\begin{align}
    \prod_{(n,N_n)\in\{\mathcal{N}\}}  \frac{(-1)^{N_n}\left(\frac{\alpha'}{2}\left(\sum_{j=2}^{i}p_j z^n_j\right)\cdot\left( \sum_{j=i+1}^{n-1}p_j z^n_j\right)\right)^{N_n}}{N_n! n^{N_n}}  \nonumber \\
     \times\prod_{(m,N_m)\in\{\widetilde{\mathcal{N}}\}}  \frac{(-1)^{N_m}\left(\frac{\alpha'}{2}\left(\sum_{j=2}^{i}p_j \bar{z}^m_j\right)\cdot\left( \sum_{j=i+1}^{n-1}p_j \bar{z}^m_j\right)\right)^{N_m}}{N_m! m^{N_m}}. \label{IN-2} 
\end{align}
According to the multinomial theorem which is
\begin{equation}
    (x_1+x_2+\ldots+x_m)^n=\sum_{\substack{k_1,k_2,\ldots,k_m\geq0\\k_1+k_2+\ldots+k_m=n}}\binom{n}{k_1,k_2,\ldots,k_m}x_1^{k_1} x_2^{k_2} \ldots x_m^{k_m}
\end{equation}
where the multinomial coefficient is
\begin{equation}
    \binom{n}{k_1,k_2,\ldots,k_m}=\frac{n!}{k_1!k_2!\ldots k_m!}
\end{equation}
Note that sum over all $k_i$ is equal to $n$. We can use this theorem to write
\begin{align}
   & \left(\frac{\alpha'}{2}\left(\sum_{j=2}^{i}p_j z^n_j\right)\cdot\left( \sum_{l=i+1}^{n-1}p_l z^n_l\right)\right)^{N_n} \nonumber \\
    &= \sum_{\substack{a^{\scriptscriptstyle{(n)}}_{jl} \geq 0 \\ \sum_{j,l} a^{\scriptscriptstyle{(n)}}_{jl} = N_n}} \frac{N_n!}{\prod_{jl} a^{\scriptscriptstyle{(n)}}_{jl}!}
    \prod_{j,l}\left( \left(\frac{\alpha'}{2}p_j \cdot p_l\right)^{a^{\scriptscriptstyle{(n)}}_{jl}} (z_j z_l)^{n a^{\scriptscriptstyle{(n)}}_{jl}} \right)
\end{align}
where $j=2,3,\ldots,i$ and $l=i+1,i+2,\ldots,n-1$. The superscript $(n)$ of the summing variables $a^{\scriptscriptstyle{(n)}}_{jl}$ is just to keep track of which $n\in\{\mathcal{N}\}$ we consider. Accordingly, the expression (\ref{IN-2}) becomes
\begin{align}
    \sum_{\substack{a^{\scriptscriptstyle{(n)}}_{jl} \geq 0 \\ \sum_{j,l} a^{\scriptscriptstyle{(n)}}_{jl} = N_n}}\prod_{n=1}^\infty\prod_{j,l}\frac{\left(-\frac{\alpha'}{2}p_j \cdot p_l\right)^{a^{\scriptscriptstyle{(n)}}_{jl}} (z_j z_l)^{n a^{\scriptscriptstyle{(n)}}_{jl}}}{a^{\scriptscriptstyle{(n)}}_{jl}! n^{a^{\scriptscriptstyle{(n)}}_{jl}}} \nonumber \\
    \times\sum_{\substack{b^{\scriptscriptstyle{(m)}}_{jl} \geq 0 \\ \sum_{j,l} b^{\scriptscriptstyle{(m)}}_{jl} = N_m}}\prod_{m=1}^\infty\prod_{j,l}\frac{\left(-\frac{\alpha'}{2}p_j \cdot p_l\right)^{b^{\scriptscriptstyle{(m)}}_{jl}} (\bar{z}_j \bar{z}_l)^{m b^{\scriptscriptstyle{(m)}}_{jl}}}{b^{\scriptscriptstyle{(m)}}_{jl}! m^{b^{\scriptscriptstyle{(m)}}_{jl}}}.
\end{align}
Note that the values of the parameters \(n\) and \(N_n\) must belong to the set \(\{\mathcal{N}\}\), as usual. Similarly, the values of \(m\) and \(N_m\) must correspond to the set \(\{\widetilde{\mathcal{N}}\}\).

We then redefine parameters such that
\begin{align}
    r_{jl}&=\sum_n a^{\scriptscriptstyle{(n)}}_{jl} \\
    s_{jl}&=\sum_m b^{\scriptscriptstyle{(m)}}_{jl} \\
    x_{jl}&=\sum_n na^{\scriptscriptstyle{(n)}}_{jl} \\
    y_{jl}&=\sum_m mb^{\scriptscriptstyle{(m)}}_{jl}
\end{align}
which are subject to
\begin{align}
    \sum_{j,l}x_{jl}=\sum_n nN_n=N \quad \text{and} \quad    \sum_{j,l}y_{jl}=\sum_m mN_m=N. \label{constraint xy}
\end{align}
Again, $N$ is a string level.

Consequently, we obtain
\begin{align}
    \sum_{x_{jl}} \sum_{r_{jl}} (-1)^N
     \prod_{j,l} \frac{s(x_{jl},r_{jl})}{x_{jl}!} \left(\frac{\alpha'}{2} p_j \cdot p_l\right)^{r_{jl}} (z_j z_l)^{x_{jl}} \nonumber \\
    \times \sum_{y_{jl}} \sum_{s_{jl}} (-1)^N
     \prod_{j,l} \frac{s(y_{jl},s_{jl})}{y_{jl}!} \left(\frac{\alpha'}{2} p_j \cdot p_l\right)^{s_{jl}} (\bar{z}_j \bar{z}_l)^{y_{jl}}
\end{align}
where the definition for the signed Stirling number of the first kind (\ref{Stirling}) was used. One can rewrite the above expression in terms of binomial coefficients using (\ref{binom}) which is
\begin{align}
    \sum_{x_{jl}} \sum_{y_{jl}}
     \prod_{j,l}\binom{\frac{\alpha'}{2}p_j\cdot p_l}{x_{jl}}\binom{\frac{\alpha'}{2}p_j\cdot p_l}{y_{jl}}(z_j z_l)^{x_{jl}} (\bar{z}_j \bar{z}_l)^{y_{jl}}. \label{IN-3}
\end{align}

After Substituting (\ref{IN-3}) to (\ref{IN-1}), one obtains 
\begin{align}
    \mathcal{I}_N=&\sum_{\{\mathcal{N}\},\{\widetilde{\mathcal{N}}\}}\int_\mathbb{C}\left( \prod_{\substack{j=2 \\j\neq i}}^{n-2} d^2z_j \right)\int\frac{d^D k}{(2\pi)^D } \sum_{x_{jl}} \sum_{y_{jl}}
     \prod_{j,l}\Bigg\{\binom{\frac{\alpha'}{2}p_j\cdot p_l}{x_{jl}}\binom{\frac{\alpha'}{2}p_j\cdot p_l}{y_{jl}}\nonumber \\
    &\times \langle0;-p_1\vert V(p_2,z_2,\bar{z}_2)\ldots V(p_{i},z_{i},\bar{z}_{i})\vert0;k\rangle (z_j z_l)^{x_{jl}} \nonumber \\
    &\times \langle0;k\vert V(p_{i+1},z_{i+1},\bar{z}_{i+1})\ldots V(p_{n-1},z_{n-1},\bar{z}_{n-1})\vert0;p_{n}\rangle(\bar{z}_j \bar{z}_l)^{y_{jl}}\Bigg\}. \label{IN-4}
\end{align}
that is subject to (\ref{constraint xy}). It is not hard to convince yourself that the above expression recovers the residue $\mathcal{B}_{i'}(x^{\scriptscriptstyle{(i')}}_{jl},y^{\scriptscriptstyle{(i')}}_{jl},\{n\},z^*_{i'})$ (\ref{Bi}) with $i=i'+1$ when tachyonic states are considered.

\section{Conclusions}
In this work, we systematically derived the BCFW recursion relations for tree-level closed string amplitudes, addressing a key challenge: the lack of an obvious structure for simple poles, unlike in open string amplitudes. To overcome this, we employed Schwinger parametrization to manipulate the Koba-Nielsen integrals, introducing $\delta$-function poles that yield simple poles upon taking residues. This allowed us to present a general expression for closed-string on-shell recursion relations, as given in (\ref{onshell-closed}).

Additionally, we performed an alternative computation of the residue for a closed string tachyon amplitude. Using the unitarity property of the amplitude, we factorized it into two fewer-point amplitudes. This was achieved by employing normalized closed string states in Fock space as a basis for constructing completeness relations. The explicit state-by-state calculation of the expression (\ref{goal2}) demonstrated consistency with previous results. Furthermore, the extension of the summation from physical states to all states in Fock space is justified by the arguments in \cite{Chang:2012qs}, which are based on the no-ghost theorem, supporting the validity of such an extension.

\section*{Acknowledgement}
This research has received funding support from the National Science, Research and Innovation Fund (NSRF) via the Program Management Unit for Human Resources \& Institutional Development, Research and Innovation [grant number B39G670016]

\bibliography{sn-bibliography}


\begin{thebibliography}{36}
\ifx \bisbn   \undefined \def \bisbn  #1{ISBN #1}\fi
\ifx \binits  \undefined \def \binits#1{#1}\fi
\ifx \bauthor  \undefined \def \bauthor#1{#1}\fi
\ifx \batitle  \undefined \def \batitle#1{#1}\fi
\ifx \bjtitle  \undefined \def \bjtitle#1{#1}\fi
\ifx \bvolume  \undefined \def \bvolume#1{\textbf{#1}}\fi
\ifx \byear  \undefined \def \byear#1{#1}\fi
\ifx \bissue  \undefined \def \bissue#1{#1}\fi
\ifx \bfpage  \undefined \def \bfpage#1{#1}\fi
\ifx \blpage  \undefined \def \blpage #1{#1}\fi
\ifx \burl  \undefined \def \burl#1{\textsf{#1}}\fi
\ifx \doiurl  \undefined \def \doiurl#1{\url{https://doi.org/#1}}\fi
\ifx \betal  \undefined \def \betal{\textit{et al.}}\fi
\ifx \binstitute  \undefined \def \binstitute#1{#1}\fi
\ifx \binstitutionaled  \undefined \def \binstitutionaled#1{#1}\fi
\ifx \bctitle  \undefined \def \bctitle#1{#1}\fi
\ifx \beditor  \undefined \def \beditor#1{#1}\fi
\ifx \bpublisher  \undefined \def \bpublisher#1{#1}\fi
\ifx \bbtitle  \undefined \def \bbtitle#1{#1}\fi
\ifx \bedition  \undefined \def \bedition#1{#1}\fi
\ifx \bseriesno  \undefined \def \bseriesno#1{#1}\fi
\ifx \blocation  \undefined \def \blocation#1{#1}\fi
\ifx \bsertitle  \undefined \def \bsertitle#1{#1}\fi
\ifx \bsnm \undefined \def \bsnm#1{#1}\fi
\ifx \bsuffix \undefined \def \bsuffix#1{#1}\fi
\ifx \bparticle \undefined \def \bparticle#1{#1}\fi
\ifx \barticle \undefined \def \barticle#1{#1}\fi
\bibcommenthead
\ifx \bconfdate \undefined \def \bconfdate #1{#1}\fi
\ifx \botherref \undefined \def \botherref #1{#1}\fi
\ifx \url \undefined \def \url#1{\textsf{#1}}\fi
\ifx \bchapter \undefined \def \bchapter#1{#1}\fi
\ifx \bbook \undefined \def \bbook#1{#1}\fi
\ifx \bcomment \undefined \def \bcomment#1{#1}\fi
\ifx \oauthor \undefined \def \oauthor#1{#1}\fi
\ifx \citeauthoryear \undefined \def \citeauthoryear#1{#1}\fi
\ifx \endbibitem  \undefined \def \endbibitem {}\fi
\ifx \bconflocation  \undefined \def \bconflocation#1{#1}\fi
\ifx \arxivurl  \undefined \def \arxivurl#1{\textsf{#1}}\fi
\csname PreBibitemsHook\endcsname

\bibitem{Neveu:1971mu}
\begin{barticle}
\bauthor{\bsnm{Neveu}, \binits{A.}},
\bauthor{\bsnm{Scherk}, \binits{J.}}:
\batitle{{Connection between Yang-Mills fields and dual models}}.
\bjtitle{Nucl. Phys. B}
\bvolume{36},
\bfpage{155}--\blpage{161}
(\byear{1972}).
\doiurl{10.1016/0550-3213(72)90301-X}
\end{barticle}
\endbibitem

\bibitem{Green:1982sw}
\begin{barticle}
\bauthor{\bsnm{Green}, \binits{M.B.}},
\bauthor{\bsnm{Schwarz}, \binits{J.H.}},
\bauthor{\bsnm{Brink}, \binits{L.}}:
\batitle{{N=4 Yang-Mills and N=8 Supergravity as Limits of String Theories}}.
\bjtitle{Nucl. Phys. B}
\bvolume{198},
\bfpage{474}--\blpage{492}
(\byear{1982}).
\doiurl{10.1016/0550-3213(82)90336-4}
\end{barticle}
\endbibitem

\bibitem{Scherk:1974ca}
\begin{barticle}
\bauthor{\bsnm{Scherk}, \binits{J.}},
\bauthor{\bsnm{Schwarz}, \binits{J.H.}}:
\batitle{{Dual Models for Nonhadrons}}.
\bjtitle{Nucl. Phys. B}
\bvolume{81},
\bfpage{118}--\blpage{144}
(\byear{1974}).
\doiurl{10.1016/0550-3213(74)90010-8}
\end{barticle}
\endbibitem

\bibitem{Yoneya:1974jg}
\begin{barticle}
\bauthor{\bsnm{Yoneya}, \binits{T.}}:
\batitle{{Connection of Dual Models to Electrodynamics and Gravidynamics}}.
\bjtitle{Prog. Theor. Phys.}
\bvolume{51},
\bfpage{1907}--\blpage{1920}
(\byear{1974}).
\doiurl{10.1143/PTP.51.1907}
\end{barticle}
\endbibitem

\bibitem{TSEYTLIN1986391}
\begin{barticle}
\bauthor{\bsnm{Tseytlin}, \binits{A.A.}}:
\batitle{Vector field effective action in the open superstring theory}.
\bjtitle{Nuclear Physics B}
\bvolume{276}(\bissue{2}),
\bfpage{391}--\blpage{428}
(\byear{1986}).
\doiurl{10.1016/0550-3213(86)90303-2}
\end{barticle}
\endbibitem

\bibitem{Gross:1986iv}
\begin{barticle}
\bauthor{\bsnm{Gross}, \binits{D.J.}},
\bauthor{\bsnm{Witten}, \binits{E.}}:
\batitle{{Superstring Modifications of Einstein's Equations}}.
\bjtitle{Nucl. Phys. B}
\bvolume{277},
\bfpage{1}
(\byear{1986}).
\doiurl{10.1016/0550-3213(86)90429-3}
\end{barticle}
\endbibitem

\bibitem{Koerber:2001uu}
\begin{barticle}
\bauthor{\bsnm{Koerber}, \binits{P.}},
\bauthor{\bsnm{Sevrin}, \binits{A.}}:
\batitle{{The NonAbelian Born-Infeld action through order alpha-prime 3}}.
\bjtitle{JHEP}
\bvolume{10},
\bfpage{003}
(\byear{2001})
{\href{https://arxiv.org/abs/hep-th/0108169}{{arXiv:hep-th/0108169}}}.
\doiurl{10.1088/1126-6708/2001/10/003}
\end{barticle}
\endbibitem

\bibitem{Metsaev:1986yb}
\begin{barticle}
\bauthor{\bsnm{Metsaev}, \binits{R.R.}},
\bauthor{\bsnm{Tseytlin}, \binits{A.A.}}:
\batitle{{Curvature Cubed Terms in String Theory Effective Actions}}.
\bjtitle{Phys. Lett. B}
\bvolume{185},
\bfpage{52}--\blpage{58}
(\byear{1987}).
\doiurl{10.1016/0370-2693(87)91527-9}
\end{barticle}
\endbibitem

\bibitem{Bergshoeff:1989de}
\begin{barticle}
\bauthor{\bsnm{Bergshoeff}, \binits{E.A.}},
\bauthor{\bparticle{de} \bsnm{Roo}, \binits{M.}}:
\batitle{{The Quartic Effective Action of the Heterotic String and Supersymmetry}}.
\bjtitle{Nucl. Phys. B}
\bvolume{328},
\bfpage{439}--\blpage{468}
(\byear{1989}).
\doiurl{10.1016/0550-3213(89)90336-2}
\end{barticle}
\endbibitem

\bibitem{Garousi:2019mca}
\begin{barticle}
\bauthor{\bsnm{Garousi}, \binits{M.R.}}:
\batitle{{Effective action of bosonic string theory at order $\alpha'^2 $}}.
\bjtitle{Eur. Phys. J. C}
\bvolume{79}(\bissue{10}),
\bfpage{827}
(\byear{2019})
{\href{https://arxiv.org/abs/1907.06500}{{arXiv:1907.06500}}}
{[hep-th]}.
\doiurl{10.1140/epjc/s10052-019-7357-4}
\end{barticle}
\endbibitem

\bibitem{Kawai:1985xq}
\begin{barticle}
\bauthor{\bsnm{Kawai}, \binits{H.}},
\bauthor{\bsnm{Lewellen}, \binits{D.C.}},
\bauthor{\bsnm{Tye}, \binits{S.H.H.}}:
\batitle{{A Relation Between Tree Amplitudes of Closed and Open Strings}}.
\bjtitle{Nucl. Phys. B}
\bvolume{269},
\bfpage{1}--\blpage{23}
(\byear{1986}).
\doiurl{10.1016/0550-3213(86)90362-7}
\end{barticle}
\endbibitem

\bibitem{Bjerrum-Bohr:2010kyi}
\begin{barticle}
\bauthor{\bsnm{Bjerrum-Bohr}, \binits{N.E.J.}},
\bauthor{\bsnm{Damgaard}, \binits{P.H.}},
\bauthor{\bsnm{Feng}, \binits{B.}},
\bauthor{\bsnm{Sondergaard}, \binits{T.}}:
\batitle{{Proof of Gravity and Yang-Mills Amplitude Relations}}.
\bjtitle{JHEP}
\bvolume{09},
\bfpage{067}
(\byear{2010})
{\href{https://arxiv.org/abs/1007.3111}{{arXiv:1007.3111}}}
{[hep-th]}.
\doiurl{10.1007/JHEP09(2010)067}
\end{barticle}
\endbibitem

\bibitem{Bjerrum-Bohr:2010mtb}
\begin{barticle}
\bauthor{\bsnm{Bjerrum-Bohr}, \binits{N.E.J.}},
\bauthor{\bsnm{Damgaard}, \binits{P.H.}},
\bauthor{\bsnm{Feng}, \binits{B.}},
\bauthor{\bsnm{Sondergaard}, \binits{T.}}:
\batitle{{New Identities among Gauge Theory Amplitudes}}.
\bjtitle{Phys. Lett. B}
\bvolume{691},
\bfpage{268}--\blpage{273}
(\byear{2010})
{\href{https://arxiv.org/abs/1006.3214}{{arXiv:1006.3214}}}
{[hep-th]}.
\doiurl{10.1016/j.physletb.2010.07.002}
\end{barticle}
\endbibitem

\bibitem{Plahte:1970wy}
\begin{barticle}
\bauthor{\bsnm{Plahte}, \binits{E.}}:
\batitle{{Symmetry properties of dual tree-graph n-point amplitudes}}.
\bjtitle{Nuovo Cim. A}
\bvolume{66},
\bfpage{713}--\blpage{733}
(\byear{1970}).
\doiurl{10.1007/BF02824716}
\end{barticle}
\endbibitem

\bibitem{Bern:2008qj}
\begin{barticle}
\bauthor{\bsnm{Bern}, \binits{Z.}},
\bauthor{\bsnm{Carrasco}, \binits{J.J.M.}},
\bauthor{\bsnm{Johansson}, \binits{H.}}:
\batitle{{New Relations for Gauge-Theory Amplitudes}}.
\bjtitle{Phys. Rev. D}
\bvolume{78},
\bfpage{085011}
(\byear{2008})
{\href{https://arxiv.org/abs/0805.3993}{{arXiv:0805.3993}}}
{[hep-ph]}.
\doiurl{10.1103/PhysRevD.78.085011}
\end{barticle}
\endbibitem

\bibitem{Kleiss:1988ne}
\begin{barticle}
\bauthor{\bsnm{Kleiss}, \binits{R.}},
\bauthor{\bsnm{Kuijf}, \binits{H.}}:
\batitle{{Multi - Gluon Cross-sections and Five Jet Production at Hadron Colliders}}.
\bjtitle{Nucl. Phys. B}
\bvolume{312},
\bfpage{616}--\blpage{644}
(\byear{1989}).
\doiurl{10.1016/0550-3213(89)90574-9}
\end{barticle}
\endbibitem

\bibitem{Bjerrum-Bohr:2009ulz}
\begin{barticle}
\bauthor{\bsnm{Bjerrum-Bohr}, \binits{N.E.J.}},
\bauthor{\bsnm{Damgaard}, \binits{P.H.}},
\bauthor{\bsnm{Vanhove}, \binits{P.}}:
\batitle{{Minimal Basis for Gauge Theory Amplitudes}}.
\bjtitle{Phys. Rev. Lett.}
\bvolume{103},
\bfpage{161602}
(\byear{2009})
{\href{https://arxiv.org/abs/0907.1425}{{arXiv:0907.1425}}}
{[hep-th]}.
\doiurl{10.1103/PhysRevLett.103.161602}
\end{barticle}
\endbibitem

\bibitem{Stieberger:2009hq}
\begin{botherref}
\oauthor{\bsnm{Stieberger}, \binits{S.}}:
{Open \& Closed vs. Pure Open String Disk Amplitudes}
(2009)
{\href{https://arxiv.org/abs/0907.2211}{{arXiv:0907.2211}}}
{[hep-th]}
\end{botherref}
\endbibitem

\bibitem{Srisangyingcharoen:2020lhx}
\begin{barticle}
\bauthor{\bsnm{Srisangyingcharoen}, \binits{P.}},
\bauthor{\bsnm{Mansfield}, \binits{P.}}:
\batitle{{Plahte Diagrams for String Scattering Amplitudes}}.
\bjtitle{JHEP}
\bvolume{04},
\bfpage{017}
(\byear{2021})
{\href{https://arxiv.org/abs/2005.01712}{{arXiv:2005.01712}}}
{[hep-th]}.
\doiurl{10.1007/JHEP04(2021)017}
\end{barticle}
\endbibitem

\bibitem{Stieberger:2023nol}
\begin{barticle}
\bauthor{\bsnm{Stieberger}, \binits{S.}}:
\batitle{{One-Loop Double Copy Relation in String Theory}}.
\bjtitle{Phys. Rev. Lett.}
\bvolume{132}(\bissue{19}),
\bfpage{191602}
(\byear{2024})
{\href{https://arxiv.org/abs/2310.07755}{{arXiv:2310.07755}}}
{[hep-th]}.
\doiurl{10.1103/PhysRevLett.132.191602}
\end{barticle}
\endbibitem

\bibitem{Yuenyong:2024ebe}
\begin{barticle}
\bauthor{\bsnm{Yuenyong}, \binits{A.}},
\bauthor{\bsnm{Srisangyingcharoen}, \binits{P.}}:
\batitle{{Relations between closed string amplitudes and mixed string amplitudes at tree-level}}.
\bjtitle{JHEP}
\bvolume{08},
\bfpage{097}
(\byear{2024})
{\href{https://arxiv.org/abs/2402.05775}{{arXiv:2402.05775}}}
{[hep-th]}.
\doiurl{10.1007/JHEP08(2024)097}
\end{barticle}
\endbibitem

\bibitem{Gerken:2020xfv}
\begin{barticle}
\bauthor{\bsnm{Gerken}, \binits{J.E.}},
\bauthor{\bsnm{Kleinschmidt}, \binits{A.}},
\bauthor{\bsnm{Mafra}, \binits{C.R.}},
\bauthor{\bsnm{Schlotterer}, \binits{O.}},
\bauthor{\bsnm{Verbeek}, \binits{B.}}:
\batitle{{Towards closed strings as single-valued open strings at genus one}}.
\bjtitle{J. Phys. A}
\bvolume{55}(\bissue{2}),
\bfpage{025401}
(\byear{2022})
{\href{https://arxiv.org/abs/2010.10558}{{arXiv:2010.10558}}}
{[hep-th]}.
\doiurl{10.1088/1751-8121/abe58b}
\end{barticle}
\endbibitem

\bibitem{Stieberger:2016lng}
\begin{barticle}
\bauthor{\bsnm{Stieberger}, \binits{S.}},
\bauthor{\bsnm{Taylor}, \binits{T.R.}}:
\batitle{{New relations for Einstein\textendash{}Yang\textendash{}Mills amplitudes}}.
\bjtitle{Nucl. Phys. B}
\bvolume{913},
\bfpage{151}--\blpage{162}
(\byear{2016})
{\href{https://arxiv.org/abs/1606.09616}{{arXiv:1606.09616}}}
{[hep-th]}.
\doiurl{10.1016/j.nuclphysb.2016.09.014}
\end{barticle}
\endbibitem

\bibitem{Dorigoni:2022npe}
\begin{barticle}
\bauthor{\bsnm{Dorigoni}, \binits{D.}},
\bauthor{\bsnm{Doroudiani}, \binits{M.}},
\bauthor{\bsnm{Drewitt}, \binits{J.}},
\bauthor{\bsnm{Hidding}, \binits{M.}},
\bauthor{\bsnm{Kleinschmidt}, \binits{A.}},
\bauthor{\bsnm{Matthes}, \binits{N.}},
\bauthor{\bsnm{Schlotterer}, \binits{O.}},
\bauthor{\bsnm{Verbeek}, \binits{B.}}:
\batitle{{Modular graph forms from equivariant iterated Eisenstein integrals}}.
\bjtitle{JHEP}
\bvolume{12},
\bfpage{162}
(\byear{2022})
{\href{https://arxiv.org/abs/2209.06772}{{arXiv:2209.06772}}}
{[hep-th]}.
\doiurl{10.1007/JHEP12(2022)162}
\end{barticle}
\endbibitem

\bibitem{Berkovits:2022ivl}
\begin{bchapter}
\bauthor{\bsnm{Berkovits}, \binits{N.}},
\bauthor{\bsnm{D'Hoker}, \binits{E.}},
\bauthor{\bsnm{Green}, \binits{M.B.}},
\bauthor{\bsnm{Johansson}, \binits{H.}},
\bauthor{\bsnm{Schlotterer}, \binits{O.}}:
\bctitle{{Snowmass White Paper: String Perturbation Theory}}.
In: \bbtitle{{Snowmass 2021}}
(\byear{2022})
\end{bchapter}
\endbibitem

\bibitem{Britto_2005}
\begin{botherref}
\oauthor{\bsnm{Britto}, \binits{R.}},
\oauthor{\bsnm{Cachazo}, \binits{F.}},
\oauthor{\bsnm{Feng}, \binits{B.}},
\oauthor{\bsnm{Witten}, \binits{E.}}:
Direct proof of the tree-level scattering amplitude recursion relation in yang-mills theory.
Physical Review Letters
\textbf{94}(18)
(2005).
\doiurl{10.1103/physrevlett.94.181602}
\end{botherref}
\endbibitem

\bibitem{Britto2005NewRR}
\begin{bchapter}
\bauthor{\bsnm{Britto}, \binits{R.}},
\bauthor{\bsnm{Cachazo}, \binits{F.}},
\bauthor{\bsnm{Feng}, \binits{B.}}:
\bctitle{New recursion relations for tree amplitudes of gluons}.
(\byear{2005})
\end{bchapter}
\endbibitem

\bibitem{Feng_2010}
\begin{botherref}
\oauthor{\bsnm{Feng}, \binits{B.}},
\oauthor{\bsnm{Wang}, \binits{J.}},
\oauthor{\bsnm{Wang}, \binits{Y.}},
\oauthor{\bsnm{Zhang}, \binits{Z.}}:
Bcfw recursion relation with nonzero boundary contribution.
Journal of High Energy Physics
\textbf{2010}(1)
(2010).
\doiurl{10.1007/jhep01(2010)019}
\end{botherref}
\endbibitem

\bibitem{RutgerBoels_2008}
\begin{barticle}
\bauthor{\bsnm{Boels}, \binits{R.}},
\bauthor{\bsnm{Larsen}, \binits{K.J.}},
\bauthor{\bsnm{Obers}, \binits{N.A.}},
\bauthor{\bsnm{Vonk}, \binits{M.}}:
\batitle{Mhv, csw and bcfw: field theory structures in string theory amplitudes}.
\bjtitle{Journal of High Energy Physics}
\bvolume{2008}(\bissue{11}),
\bfpage{015}
(\byear{2008}).
\doiurl{10.1088/1126-6708/2008/11/015}
\end{barticle}
\endbibitem

\bibitem{Boels:2010bv}
\begin{barticle}
\bauthor{\bsnm{Boels}, \binits{R.H.}},
\bauthor{\bsnm{Marmiroli}, \binits{D.}},
\bauthor{\bsnm{Obers}, \binits{N.A.}}:
\batitle{{On-shell Recursion in String Theory}}.
\bjtitle{JHEP}
\bvolume{10},
\bfpage{034}
(\byear{2010})
{\href{https://arxiv.org/abs/1002.5029}{{arXiv:1002.5029}}}
{[hep-th]}.
\doiurl{10.1007/JHEP10(2010)034}
\end{barticle}
\endbibitem

\bibitem{Chang:2012qs}
\begin{barticle}
\bauthor{\bsnm{Chang}, \binits{Y.-Y.}},
\bauthor{\bsnm{Feng}, \binits{B.}},
\bauthor{\bsnm{Fu}, \binits{C.-H.}},
\bauthor{\bsnm{Lee}, \binits{J.-C.}},
\bauthor{\bsnm{Wang}, \binits{Y.}},
\bauthor{\bsnm{Yang}, \binits{Y.}}:
\batitle{{A note on on-shell recursion relation of string amplitudes}}.
\bjtitle{JHEP}
\bvolume{02},
\bfpage{028}
(\byear{2013})
{\href{https://arxiv.org/abs/1210.1776}{{arXiv:1210.1776}}}
{[hep-th]}.
\doiurl{10.1007/JHEP02(2013)028}
\end{barticle}
\endbibitem

\bibitem{Cheung:2010vn}
\begin{barticle}
\bauthor{\bsnm{Cheung}, \binits{C.}},
\bauthor{\bsnm{O'Connell}, \binits{D.}},
\bauthor{\bsnm{Wecht}, \binits{B.}}:
\batitle{{BCFW Recursion Relations and String Theory}}.
\bjtitle{JHEP}
\bvolume{09},
\bfpage{052}
(\byear{2010})
{\href{https://arxiv.org/abs/1002.4674}{{arXiv:1002.4674}}}
{[hep-th]}.
\doiurl{10.1007/JHEP09(2010)052}
\end{barticle}
\endbibitem

\bibitem{Srisangyingcharoen:2024qyx}
\begin{barticle}
\bauthor{\bsnm{Srisangyingcharoen}, \binits{P.}}:
\batitle{{General expressions for on-shell recursion relations for tree-level open string amplitudes}}.
\bjtitle{Phys. Lett. B}
\bvolume{858},
\bfpage{139038}
(\byear{2024})
{\href{https://arxiv.org/abs/2404.00244}{{arXiv:2404.00244}}}
{[hep-th]}.
\doiurl{10.1016/j.physletb.2024.139038}
\end{barticle}
\endbibitem

\bibitem{lepowsky2012introduction}
\begin{bbook}
\bauthor{\bsnm{Lepowsky}, \binits{J.}},
\bauthor{\bsnm{Li}, \binits{H.}}:
\bbtitle{Introduction to Vertex Operator Algebras and Their Representations}.
\bsertitle{Progress in Mathematics}.
\bpublisher{Birkh{\"a}user Boston}, \blocation{???}
(\byear{2012}).
\burl{https://books.google.co.th/books?id=bB3UBwAAQBAJ}
\end{bbook}
\endbibitem

\bibitem{Virasoro:1969me}
\begin{barticle}
\bauthor{\bsnm{Virasoro}, \binits{M.A.}}:
\batitle{{Alternative constructions of crossing-symmetric amplitudes with regge behavior}}.
\bjtitle{Phys. Rev.}
\bvolume{177},
\bfpage{2309}--\blpage{2311}
(\byear{1969}).
\doiurl{10.1103/PhysRev.177.2309}
\end{barticle}
\endbibitem

\bibitem{Shapiro:1970gy}
\begin{barticle}
\bauthor{\bsnm{Shapiro}, \binits{J.A.}}:
\batitle{{Electrostatic analog for the virasoro model}}.
\bjtitle{Phys. Lett. B}
\bvolume{33},
\bfpage{361}--\blpage{362}
(\byear{1970}).
\doiurl{10.1016/0370-2693(70)90255-8}
\end{barticle}
\endbibitem

\end{thebibliography}


\end{document}